\documentclass{siamltex}

\usepackage{amssymb}
\usepackage{amsmath}
\usepackage[left=2cm,right=2cm,top=2cm,bottom=2cm]{geometry}

\usepackage{color}

\usepackage{graphicx}
\usepackage{subcaption}

\usepackage{enumitem}

\usepackage{float}

\newcommand{\bq}{{\bf q}}
\newcommand{\bp}{{\bf p}}
\newcommand{\bb}{{\bf b}}
\newcommand{\eps}{\varepsilon}

\title{A Discrete-to-Continuum Model of Weakly Interacting
  Incommensurate Chains}

\author{
    Malena I.~Espa\~ nol\footnotemark[1]
\and
    Dmitry Golovaty\footnotemark[1]
\and
    J. Patrick Wilber\footnotemark[1]}

\begin{document}
\maketitle

\renewcommand{\thefootnote}{\fnsymbol{footnote}}
\footnotetext[1]{
    Department of Mathematics,
    The University of Akron,
    Akron,
    OH 44325, USA.}

\begin{abstract}
In this paper we use a formal discrete-to-continuum procedure to derive a continuum variational model for two chains of atoms with slightly incommensurate lattices. The chains represent a cross-section of a three-dimensional system consisting of a graphene sheet suspended over a substrate. The continuum model recovers both qualitatively and quantitatively the behavior observed in the corresponding discrete model. The numerical solutions for both models demonstrate the presence of large commensurate regions separated by localized incommensurate domain walls. 
\end{abstract}

\begin{keywords} supported graphene, moir\'e patterns,
  discrete-to-continuum modeling\end{keywords}

\section{Introduction} \label{s1}

A graphene sheet is a single-atom thick sheet of carbon atoms
arranged in a hexagonal lattice.  Graphene has exceptional physical
properties and yields insights into the fundamental physics of
two-dimensional materials.  These facts have motivated an extensive
effort to model and simulate graphene and other related carbon
nanostructures.

The study in this paper is motivated by discrete-to-continuum modeling that describes
the deformation of a graphene sheet suspended over a substrate.  For a
suspended graphene sheet, the positions of the atoms on the sheet are
determined by the strong, bonded interactions between nearest
neighbors on the sheet and by the weak, non-bonded interactions with
nearby atoms on the substrate.  A mismatch between the geometries of
the hexagonal lattice of graphene and the substrate lattice can induce
strain in the sheet.  This strain can be relaxed by both in-plane and
out-of-plane, atomic-scale displacements of the atoms on the sheet.

Insight into the response of graphene to this lattice
mismatch can be obtained within the framework of the classical
Frenkel-Kontorova theory \cite{braun2013frenkel}. 
For slightly mismatched one-dimensional lattices, this theory predicts
relatively large commensurate regions separated by localized
incommensurate regions.  Hence we expect for suspended graphene that
the local adjustments of atoms on the sheet may create large domains
where the two lattices are commensurate and the interlayer energy is
minimized.  At the same time, there should be localized incommensurate
regions, or domain walls, where strain may be relaxed by out-of-plane
displacement.

Commensurate regions separated by domain walls have been observed in
simulations of relaxed moir\'e patterns in graphene sheets
\cite{van2015relaxation,van2014moire}.  When two lattices with
different lattice geometries or the same geometry but different
orientations are stacked, a larger periodic pattern, called a moir\'e
pattern, emerges.  These moir\'e patterns are a strictly visual effect.
However, if the atoms on one or both of the lattices are then relaxed
to accommodate the mismatch between the lattices, additional patterns
can occur.  In \cite{van2015relaxation}, the authors study these
relaxed moir\'e patterns by simulating interacting, identical graphene
lattices where one lattice is slightly rotated with respect to the
other.  In \cite{van2014moire}, the authors report on similar
simulations for a slightly rotated graphene lattice interacting with a
hexagonal boron nitride substrate, which also has the structure of a
hexagonal lattice with a slightly larger lattice constant than that of
graphene.  In both papers, simulations predict a two-dimensional pattern {
of wrinkles intersecting at small, misfit regions, called hot spots, exhibiting large
out-of-plane displacements}. The {wrinkles} separate large, flat
domains of commensurate regions.

For the model we develop in this paper, the starting point is a pair
of parallel curves that `carry' atoms. The lower curve models a rigid
substrate, and its atoms are fixed at a prescribed spacing.
The upper curve models a graphene sheet, and the
atoms on the upper curve can displace.  The equilibrium spacing
between the atoms on the upper curve is also prescribed.  The two
lattices are mismatched if the interatomic spacing between the atoms on the lower curve
does not equal the equilibrium spacing between atoms on the upper
curve.

We start with a discrete description of the stretching and bending
energies of the atoms on the upper curve and the weak interaction
energy between the atoms on the upper and lower curves, respectively.  From
this we attain a continuum description when a small parameter,
representing the ratio of the typical spacing between the curves to
the length of the curves, is small but not equal to zero.  The minimizers of this
continuum energy represent equilibrium configurations of the upper
curve.  

The continuum energy that we obtain has a Ginzburg-Landau-type
structure with the elastic contribution that corresponds to the
classical F\"oppl-von K\'arm\'an energy \cite{PhysRevE.85.066115}. A
number of recent studies have considered related problems of wrinkling
of thin elastic sheets that are, e.g., bonded to a compliant substrate
with a large compressive misfit \cite{Kohn2013,wang2014phase,PhysRevLett.72.3570}, pulled down by the
force of gravity \cite{CPA:CPA21643}, or are floating on a fluid
\cite{PhysRevLett.107.044301,PhysRevLett.105.038302,CPA:CPA21471}.  

The novel aspect of the continuum energy derived in the present work is the {potential} describing
the weak interactions between the curves.  Although a continuum
description, the weak energy retains information about the {local} mismatch
between the original discrete lattices. {The presence of lattice mismatch constitutes the principal difference between the adhesion forces that are associated with the weak interactions between the lattices and the forces that are usually considered in the standard problem for a thin film bonded to an adhesive substrate. The wrinkling patterns in a typical adhesion problem develop primarily due to the misfit strains. On the other hand, the wrinkles observed in incommensurate lattices appear not only because of the mismatch in strains, but also in order to accommodate the difference between the number of atoms on the two chains while maximizing the regions of registry. For example, it does not make a difference whether the deformable chain has one more or one fewer atom than the rigid chain per period---in both cases, a single wrinkle will appear wherever there is an extra atom or a "vacancy" on the deformable chain. The out-of-plane deflection at this position occurs because the equilibrium distances between the chains are different if the chains are in registry or out of registry.}

From our total continuum energy we derive the Euler-Lagrange
equations, which are then solved numerically.  We present some basic
comparisons between discrete simulations and the predictions of our
continuum model.  {For curves corresponding to slightly mismatched chains}, our model predicts
large commensurate regions separated by domain walls, or wrinkles, formed by
localized out-of-plane ridges.  The spacing of these
domain walls can be determined by the need to accommodate a certain
number of extra atoms on the upper curve.  Qualitatively, our
solutions exhibit a pattern of {wrinkles} similar
to the predictions of the atomistic simulations in
\cite{van2015relaxation}.

The discrete-to-continuum modeling in this paper is analogous to the
approach taken in \cite{golovaty2008continuum}, in which the authors
derive a continuum theory of multi-walled carbon nanotubes by
upscaling an atomistic model.  The atomistic model includes a bending
energy related to strong covalent bonds between atoms in the same wall
and an interaction energy between the atoms in adjacent walls.
Numerical solutions using the continuum model that results from
upscaling show that, for sufficiently large radii, the cross-section
of a double-walled nanotube polygonizes.  In this setting, the
straight sections of the polygonal cross-section are commensurate
regions.  The corners of the polygon are the domain walls, and are
analogous to the localized ridges predicted by the model developed in
this paper.

A discrete model, similar to the one considered in this paper, was recently used in
\cite{cazeaux2017analysis} to demonstrate numerically that the
spontaneous atomic-scale relaxation of free-standing systems of
incommensurate van der Waals bilayers leads to a simultaneous
long-range rippling of the bilayer system.

This paper is organized as follows. In Section 2, we formulate a
discrete energy of the system of a graphene sheet over a substrate.
In Section 3, we derive a continuum energy that keeps
track of the mismatch of the spacing between the atoms on each
curve.  The next section includes numerical results that compare the
atomistic model with the continuum model. Furthermore, in this section
we show how the different parameters give rise to different material
deformations.  

\section{Atomistic Model} \label{s5}

For simplicity, here we model the formation of isolated wrinkles in a graphene layer
supported by a substrate within the framework of a one-dimensional model. Note that a similar description can be readily developed for graphene bilayers (cf. \cite{golovaty2008continuum}).   

Suppose that we have a discrete system that consists of
two chains of atoms $\hat{\mathcal C}_1$ and $\hat{\mathcal C}_2$ that are $L$
units long. The atoms on the bottom chain $\hat{\mathcal C}_1$ are spaced
$h_1$ apart and cannot move. This chain describes a rigid
substrate. The atoms on the top chain $\hat{\mathcal C}_2$ can move. This 
chain represents a deformable graphene layer that is nearly
inextensible and has a finite resistance to bending.  Each atom on the
top chain interacts with its two nearest neighbors via a strong
bond potential, represented here by a stiff linear spring such that
the equilibrium spacing between the atoms on $\hat{\mathcal C}_2$ is
$h_2$. The resistance to bending is modeled by torsional springs
between adjacent bonds. Further, all atoms on the first chain are
assumed to interact with all atoms on the second chain via interatomic
van der Waals potential. In what follows we will refer to $\hat{\mathcal C}_1$ as the rigid chain and to $\hat{\mathcal C}_2$ as the deformable chain.

We assume that, in the reference configuration (Figure~\ref{f1}), the chains are
parallel and separated by a distance $\sigma$. Here $\sigma$ is equal
to the equilibrium distance between two atoms interacting via the van der Waals forces. Note that this reference configuration is {\em not} in equilibrium. If the ratio between $\sigma$ and the equilibrium bond length $h_1$ is large enough, from the point of view of an atom on $\hat{\mathcal C}_2$, its van der Waals interaction with all atoms on $\hat{\mathcal C}_1$ can be represented by an interaction with the curve representing $\hat{\mathcal C}_1$ with a uniform
atomic density \cite{wilber2007continuum}. In equilibrium, the curves  $\hat{\mathcal C}_1$ and $\hat{\mathcal C}_2$ are then given by the two straight parallel lines.  The distance between these lines should be slightly smaller than $\sigma$ in order to accommodate attractive forces from more distant atoms. This approximation, however, ignores possible registry
effects that are significant in determining the shape of the
deformable chain $\hat{\mathcal C}_2$. In fact, the only situation in which
the two straight parallel chains {\it would} correspond to an equilibrium
configuration is when $h_1=h_2$. Indeed, in this case, all atoms on
$\hat{\mathcal C}_2$ would occupy the positions above the midpoints between
the atoms on $\hat{\mathcal C}_1$ and the system would be in {\it global
  registry}.

Here we are concerned with the situation when $h_1\neq h_2$, but
$|h_1-h_2|/h_1\ll1$. Under these assumptions, global registry cannot
be achieved in an undeformed configuration because the distance
between midpoints of neighboring intervals in $\hat{\mathcal C}_1$ is not
equal to $h_2$. It follows that in order to achieve equilibrium, the
deformable chain would have to adjust by some combination of bending
and stretching.

Let the current positions of the $N_2$ atoms on the chain be given by the vectors $\left\{\bq_1,\dots, \bq_{N_2}\right\}\subset \mathbb{R}^{2}$. For every $i=1,\ldots,N_2-1$, we represent the bond between the atom $i$ and the atom $i+1$ by the vector $\bb_i = \bq_{i+1}-\bq_i$. Then, the total energy of the system is given by 
$$E({\bq_1,\ldots,\bq_{N_2}}) = E_s({\bq_1,\ldots,\bq_{N_2}})  + E_b({\bq_1,\ldots,\bq_{N_2}})  + E_w({\bq_1,\ldots,\bq_{N_2}}) ,$$
where the stretching energy is defined by a harmonic potential
\begin{equation}\label{eq:DiscreteE_s}
E_s({\bq_1,\ldots,\bq_{N_2}})  = \sum_{i=1}^{N_2-1} \frac{k_s}{2}\left( \frac{\|\bb_i\|-h_2}{h_2}\right)^2, 
\end{equation}
with $k_s$ being the spring constant. The bending between the adjacent links of the chain is penalized by introducing torsional springs connecting these links, and therefore, the associated bending energy is given by
\begin{equation}
  E_{b}({\bq_1,\ldots,\bq_{N_2}}) 
  =
  \sum_{i=1}^{N_2-2} \frac{k_b}{2}
  \left(\theta_{i+1}-\theta_i\right)^2, \label{e21}
\end{equation}
where $k_b$ is the torsional constant and $\theta_i$ is the angle between the $i$-th link and the $x$-axis defined by 
\[\bq_{i+1} - \bq_{i} =  \|\bb_{i}\| (\cos(\theta_i), \sin (\theta_i)),\] 
for every $i=1,\ldots,N_2-1$. Assuming that $\left|\theta_{i+1}-\theta_i\right|\ll1$ for all $i=1,\ldots,N_2-1$, in the sequel we will consider the expression 
\begin{equation}
  E_{b}({\bq_1,\ldots,\bq_{N_2}}) 
  =
  k_b\sum_{i=2}^{N_2-1} 
  \frac{\|\bb_i\|\|\bb_{i-1}\|-\bb_i\cdot\bb_{i-1}}{\|\bb_{i}\|\|\bb_{i-1}\|}, \label{e21.5}
\end{equation}
 for the bending energy that is equivalent to \eqref{e21} to leading order.

The energy of the weak van der Waals interaction between $\hat{\mathcal C}_1$ and $\hat{\mathcal C}_2$ is defined by
\begin{equation}\label{eq:DiscreteE_w}
  E_w({\bq_1,\ldots,\bq_{N_2}}) 
  =
  \omega\sum_{i=1}^{N_2} \sum_{j=1}^{N_1} g\left(\frac{\|\bp_j - \bq_i \|}{\sigma}\right),
\end{equation}  
where $g$ is a given weak pairwise interaction potential and $\bp_j=(p_j,0)\in\mathbb R^2,\ j=1,\ldots,N_1$ are the positions of the atoms on the rigid chain.  The parameters $\sigma$ and $\omega$ define the equilibrium interatomic distance and the strength of the potential energy \eqref{eq:DiscreteE_w}, respectively. In what follows, we assume that $g$ is the classical Lennard-Jones 12-6 potential given by
$$g(r) = r^{-12}-2r^{-6}.$$ 

\section{Continuum Model} \label{s2}

\begin{figure}[htb]
\hspace*{.1\textwidth}
  \begin{subfigure}[b]{0.8\textwidth}
    \includegraphics[width=.9\linewidth, clip, trim=1.2in 1.4in 0in 1in]
                    {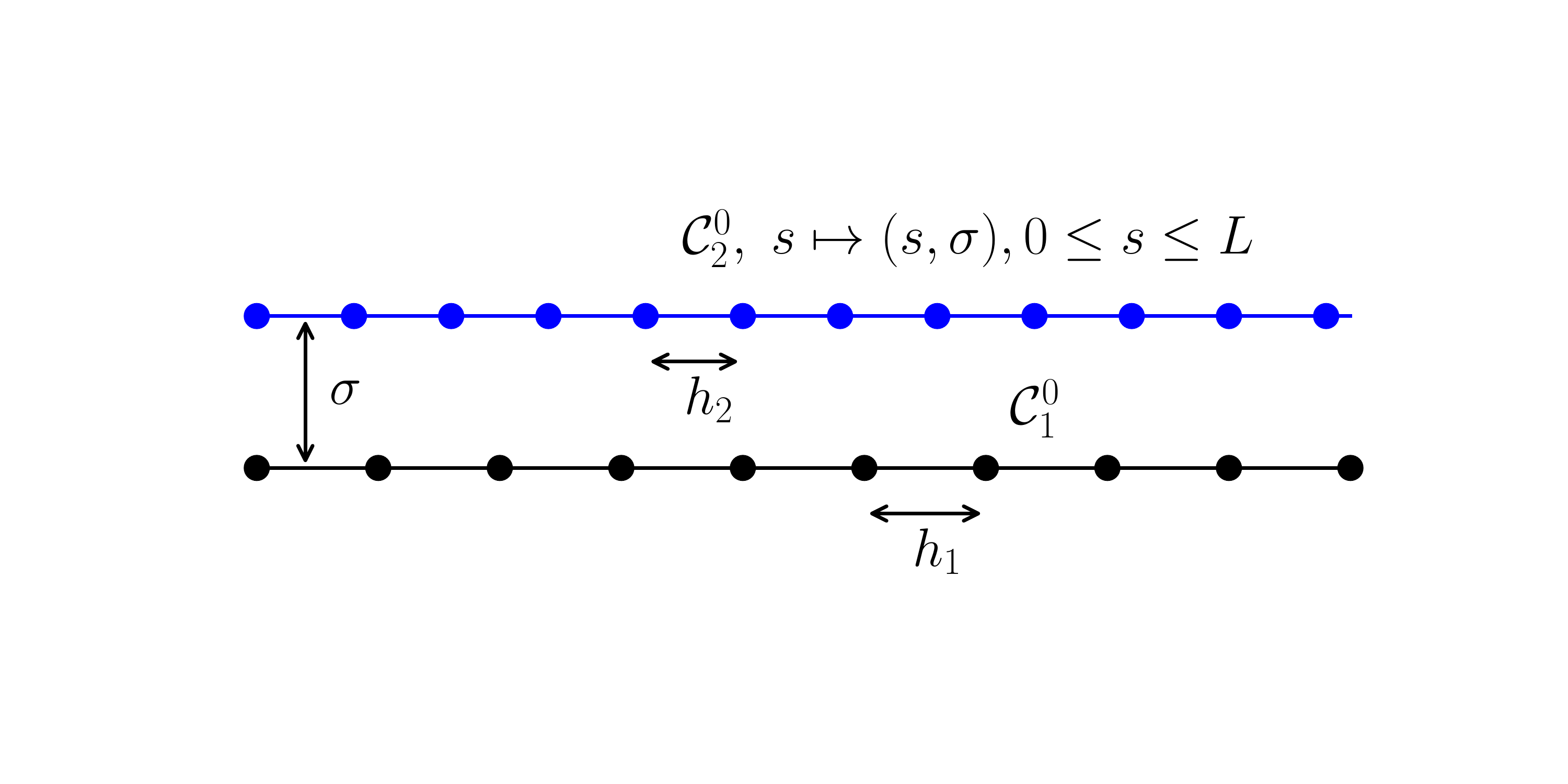}
    \caption{Reference Configuration}
    \label{f1}
  \end{subfigure}

\hspace*{.1\textwidth}
  \begin{subfigure}[b]{0.8\textwidth}
    \includegraphics[width=.9\linewidth, clip, trim=1.2in 2in 0in 1in]
                    {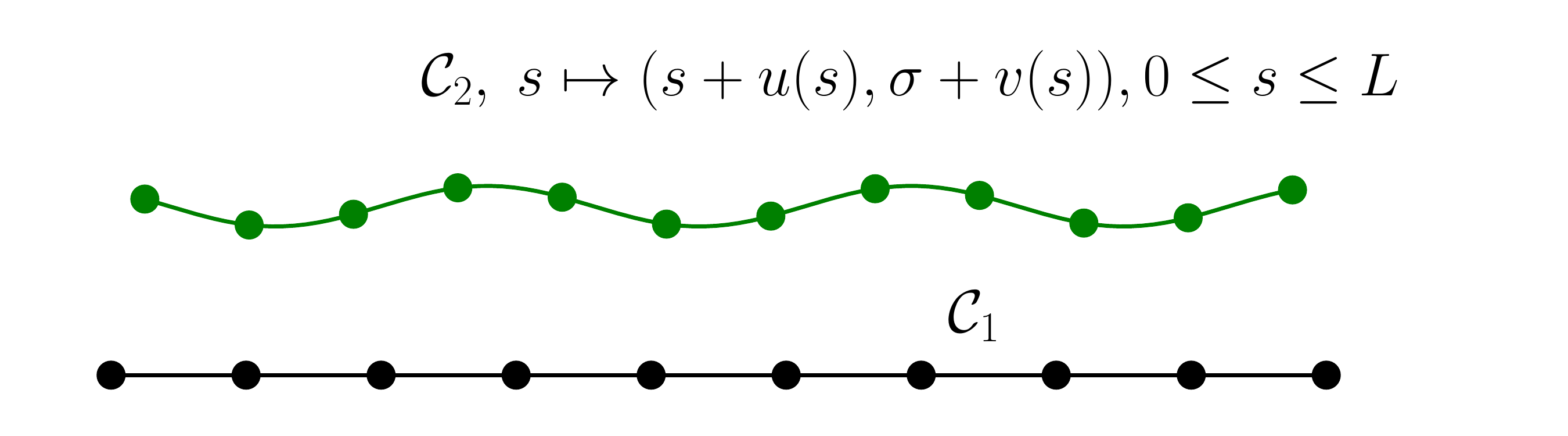}
    \caption{Deformed Configuration}
    \label{f2}
  \end{subfigure}
  \caption{The reference (a) and the deformed (b) configurations of the system of two chains in dimensional coordinates}\label{f11}
\end{figure}

As a first step in formally deriving a continuum model, we assume that the chains of atoms
are embedded in two sufficiently smooth curves $\mathcal C_1$ and $\mathcal
C_2$.  See Figure~\ref{f1}.  Hence, the lower curve $\mathcal C_1$ is
straight and rigid, while the upper curve $\mathcal C_2$ can deform.
We denote these two curves in the reference configuration by 
$\mathcal C^{0}_1$ and $\mathcal C^{0}_2$.
We assume
\begin{equation}
  \mathcal C^{0}_1 = \{(s,0):s\in[0,L]\}
  \mbox{ and }
  \mathcal C^{0}_2 = \left\{(s,\sigma):s\in[0,L]\right\}.
  \label{ee14}
\end{equation}
Letting $h_{i},\ i=1,2,$ denote the equilibrium spacing for the atoms on
$\mathcal C^{0}_i$, we assume that the atoms on $\mathcal C^{0}_i$ are
a distance $h_{i}$ apart in the reference configuration. Note, that we do not assume that the system of two chains is stress-free in the reference configuration.

We let $(u(s),v(s))$ be the displacement of the point $(s,\sigma)$
on $\mathcal C^{0}_2$.  Hence the deformed
curve $\mathcal C_2$ is given by
\begin{equation}
  \left\{(s+u(s),\sigma+v(s)):s\in[0,L]\right\}. 	
  \label{ee12}
\end{equation}
In particular, an atom at the point $(s_{i},\sigma)$
on $\mathcal C^{0}_2$ is displaced to the point
$(s_{i}+u(s_{i}),\sigma+v(s_{i}))$.
See Figure~\ref{f2}.

We assume $\sigma<<L$, i.e., that the spacing between the chains is much
less than the length of the chains.  To exploit this, we introduce the rescalings
\begin{equation}
   t = \frac{s}{L},\ \ 
  \bar{\xi} = \frac{u}{L},\ \ 
  \bar\eta = \frac{v}{L}, \ \
  \bar E=\frac{E}{\omega} 
  \label{ee1}
\end{equation}
and the nondimensional parameters
\begin{equation}
  \varepsilon = \frac{\sigma}{L},\ \ 
  \delta_{1} = \frac{h_{1}}{\sigma},\ \ 
  \delta_{2} = \frac{h_{2}}{\sigma},\ \ 
  \gamma_e=\frac{k_s}{\omega\delta_2}, \ \
  \gamma_b=\frac{k_b\delta_2}{\omega}.
  \label{ee1.1}
\end{equation}
We obtain with a slight abuse of notation that
\begin{equation}
  \mathcal C_1
  =
  \left\{(t,0)\colon t\in[0,1]\right\}
 \mbox{ and }
  \mathcal C_2
  =
  \left\{(t+\bar\xi(t),\varepsilon+\bar \eta(t))\colon t\in[0,1]\right\}.
  \label{ee13}
\end{equation}
We assume that $\delta_i=\mathcal{O}(1),\ i=1,2$,
i.e., the distance between the atoms within each chain is comparable to the distance
between the chains (and hence both are much smaller than the length of the
chains).  Furthermore, in order to observe the registry effects on a macroscale,
we assume that
\begin{equation}
  \delta_{1}-\delta_{2}=\mathcal{O}(\varepsilon)
  \label{ee9},
\end{equation}
so that the mismatch between the equilibrium spacings of the two chains is small.

\begin{figure}[htb]
\hspace*{.1\textwidth}
  \begin{subfigure}[b]{0.8\textwidth}
    \includegraphics[width=.9\linewidth, clip, trim=1.2in 1.4in 0in 1in]
                    {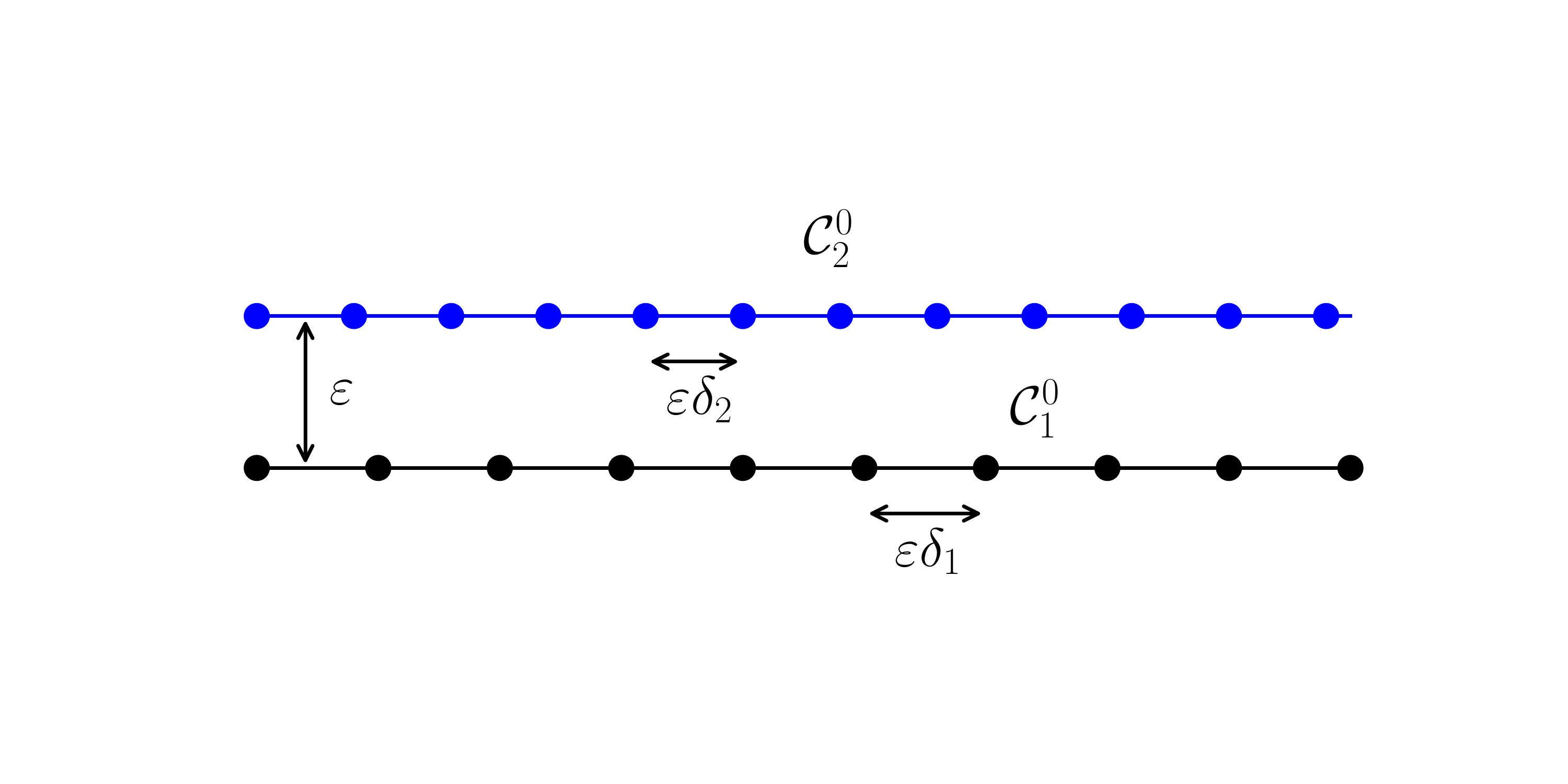}
    \caption{Reference Configuration}
    \label{f3}
  \end{subfigure}

\hspace*{.1\textwidth}  
  \begin{subfigure}[b]{0.8\textwidth}
    \includegraphics[width=.9\linewidth, clip, trim=1.2in 2in 0in 1in]
                    {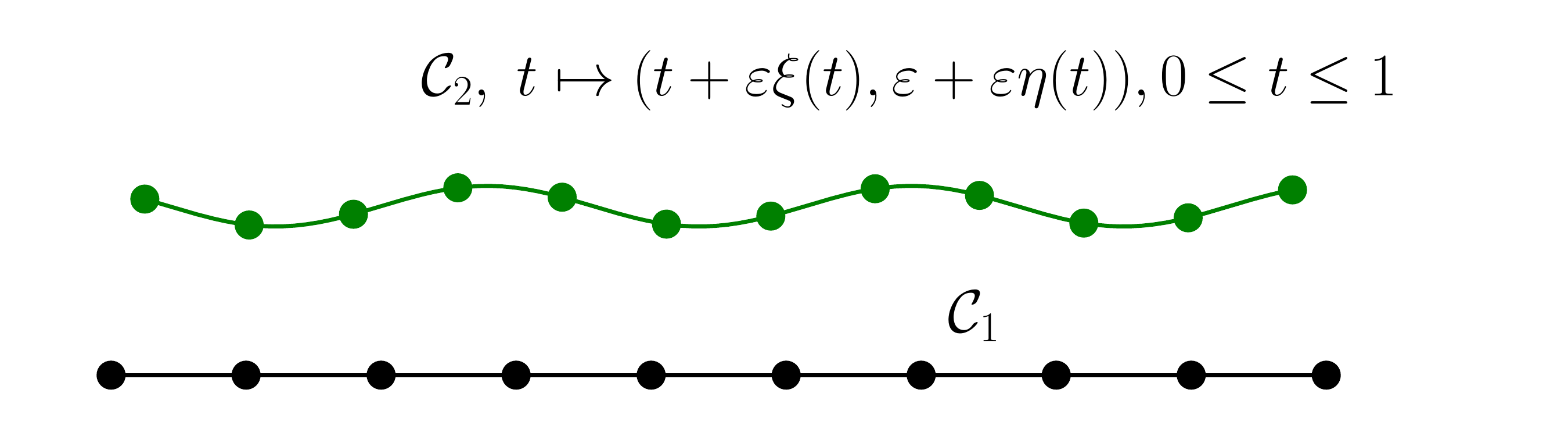}
    \caption{Deformed Configuration}
    \label{f4}
  \end{subfigure}
    \caption{The reference (a) and the deformed (b) configurations of the system of two chains in nondimensional coordinates}\label{ff35}

\end{figure}

We assume $\bar{\xi}=\varepsilon\xi$, $\bar{\eta}=\varepsilon\eta$,
so that $\mathcal{C}_{2}$ is parameterized by
\begin{equation}
 \Gamma(t)
  = (t+\varepsilon\xi(t), \varepsilon+\varepsilon \eta(t)) \mbox{ with } t\in[0,1].
  \label{ee10}
\end{equation}
See Figures~\ref{f3} and \ref{f4}.  
These scalings for the displacements are appropriate for small deformations considered here and
eventually will lead to expressions for the strains similar to those for F\"oppl-von
K\'arm\'an theory.  

In the rescaled coordinates, the atoms in $\mathcal{C}_{2}^0$ are located at the points $\bq_i^0=(t_{i},\eps)$, were $t_i=\eps\delta_2i$ for $i=1,\ldots,N_2$. The $i$-th atom is then displaced to the point
$\bq_i=(t_{i}+\eps\xi(t_{i}),\eps+\eps\eta(t_{i}))$ for every $i=1,\ldots,N_2$. 

\subsection{Elastic Energy Contribution} \label{elencon}

We now take advantage of the fact that $\eps\ll1$. By using the Taylor expansions of $\xi(t_i)$ and $\eta(t_i)$ in $\eps$, the bond $\bb_i$ between the atom $i$ and the atom $i+1$ can be expressed in the form
\begin{multline}
\label{eq:bond}
\bb_i =\eps(\delta_2+\xi(t_{i+1})-\xi(t_{i}),\eta(t_{i+1})-\eta(t_{i}))\\=\eps\delta_2\left(1+\eps\xi^\prime(t_i)+\frac{\delta_2\eps^2}{2}\xi^{\prime\prime}(t_i)+\frac{\delta_2^2\eps^3}{6}\xi^{\prime\prime\prime}(t_i),\eps\eta^\prime(t_i)+\frac{\delta_2\eps^2}{2}\eta^{\prime\prime}(t_i)+\frac{\delta_2^2\eps^3}{6}\eta^{\prime\prime\prime}(t_i)\right)+o\left(\eps^3\right).
\end{multline}
Substituting these expansions into the expressions \eqref{eq:DiscreteE_s} and \eqref{e21.5} for the stretching and bending energies, respectively, we can redefine both energies in terms of values of $\xi$ and $\eta$ at $t_i$, where $i=1,\ldots,N_2-1$. Preserving the notation for the energies, we have that
\begin{multline}
\label{stretch}
\bar E_s[\xi,\eta]=\frac{\gamma_e\delta_2\eps^2}{2}\sum_{i=1}^{N_2-1} \left\{{\left[\xi^\prime(t_i)+\frac{\eps}{2}\left({\left(\eta^\prime(t_i)\right)}^2+\delta_2\xi^{\prime\prime}(t_i)\right)\right]}^2\right.\\\left.-\frac{\eps^2}{3}\left[3{\left(\xi^\prime(t_i)\eta^\prime(t_i)\right)}^2-3\delta_2\xi^\prime(t_i)\eta^\prime(t_i)\eta^{\prime\prime}(t_i)-\delta_2^2\xi^\prime(t_i)\xi^{\prime\prime\prime}(t_i)\right]\right\}+o\left(N_2\eps^4\right),
\end{multline}
and
\begin{equation}
\label{bend}
\bar E_b[\xi,\eta]= \frac{\gamma_b\delta_2\eps^4}{2}\sum_{i=2}^{N_2-1} {\left[\eta^{\prime\prime}(t_i)\right]}^2+o\left(N_2\eps^4\right).
\end{equation}
Here the number of terms in the expansion of $\bar E_s$ is chosen so as to match the power of $\eps$ in the lowest order term in the expansion of $\bar E_b$.

We now recall that $\mathcal C_2^0$ has a unit length in nondimensional coordinates and that the spacing between the atoms is equal to $\eps\delta_2\ll1$. Hence, the number of atoms in $\mathcal C_2^0$ is approximately $\frac1\eps$ and therefore 
\begin{multline}
\label{stretch_cont_0}
\bar E_s[\xi,\eta]\sim\frac{\gamma_e\eps}{2}\int_0^1 \left\{{\left[\xi^\prime(t)+\frac{\eps}{2}\left({\left(\eta^\prime(t)\right)}^2+\delta_2\xi^{\prime\prime}(t)\right)\right]}^2\right.\\\left.-\frac{\eps^2}{3}\left[3{\left(\xi^\prime(t)\eta^\prime(t)\right)}^2-3\delta_2\xi^\prime(t)\eta^\prime(t)\eta^{\prime\prime}(t)-\delta_2^2\xi^\prime(t)\xi^{\prime\prime\prime}(t)\right]\right\}\,dt=:\mathcal E_s^\eps[\xi,\eta]
\end{multline}
and
\begin{equation}
\label{bend_cont}
\bar E_b[\xi,\eta]\sim\frac{\gamma_b\eps^3}{2}\int_0^1 {\left[\eta^{\prime\prime}(t)\right]}^2\,dt=:\mathcal E_b^\eps[\xi,\eta].
\end{equation}

From now on, we assume that the admissible functions $\xi$ and $\eta$ satisfy periodic boundary conditions, i.e., $\xi(0)=\xi(1)$, $\eta(0)=\eta(1)$, $\xi^\prime(0)=\xi^\prime(1)$, $\eta^\prime(0)=\eta^\prime(1)$, etc. With this assumption, after integrating by parts, the stretching energy in \eqref{stretch_cont_0} can be written as 
\begin{equation}
\label{stretch_cont}
\mathcal E_s^\eps[\xi,\eta]=\frac{\gamma_e\eps}{2}\int_0^1 \left\{{\left(\xi^\prime(t)+\frac{\eps}{2}{\left(\eta^\prime(t)\right)}^2\right)}^2-\eps^2{\left(\xi^\prime(t)\eta^\prime(t)\right)}^2-\frac{\eps^2\delta_2^2}{12}{\left(\xi^{\prime\prime}(t)\right)}^2\right\}\,dt.
\end{equation}
{Here the term ${\left(\xi^\prime(t)+\frac{\eps}{2}{\left(\eta^\prime(t)\right)}^2\right)}^2$ is a one-dimensional version of the F\"oppl-von K\'arm\'an energy describing large deflections of thin flat plates. Note that the remaining higher-order nonlinear elastic terms coupling the vertical and horizontal components of deformation are nonpositive and thus problematic from the point of view of establishing existence of a solution to the corresponding variational problem. In Section \ref{encon} we argue that neglecting these terms should not affect the global behavior of a minimizer of the full continuum energy.}

 \subsection{Van der Waals Energy Contribution} \label{vaencon}

We now discuss the contribution to the energy from the van der Waals
interactions, that is, the continuum version of \eqref{eq:DiscreteE_w}, which has the form of
\begin{equation}
  \mathcal E^\eps_w[\xi,\eta]=\frac{1}{\varepsilon}\int_0^1 G\left(t,\xi,\eta\right)\,dt.
\end{equation}
The novelty of our model is in defining a function $G$ that gives a
continuum description of the mismatch of the spacing between the atoms
on each chain. We shall first focus in the inner sum of \eqref{eq:DiscreteE_w} and try to estimate the interaction of a given atom on the deformable chain with all the atoms on the rigid chain.  We accomplish this by tracking the offset between the two lattices embedded in $\mathcal C_1$ and $\mathcal C_2$ as a function of $t$. 

Our starting point is to pick an atom $i$ on $\mathcal C_2$. A
discrete description of the total interaction energy between this atom and the atoms on $\mathcal{C}_{1}$ is given by
\begin{equation}
  \sum_{j=-\infty}^{\infty}g(d_{ij}/\varepsilon),
  \label{ee3}
\end{equation}
where $d_{ij}$ is the distance between the fixed atom $i$ on $\mathcal{C}_{2}$ and the atom $j$ on $\mathcal{C}_{1}$. Here we replace the finite rigid chain with a larger rigid chain that contains infinitely many atoms. Due to the rapid decay of the potential function $g$ as it argument tends to $\infty$, this should have no effect on the total interaction energy unless the atom $i$ lies very close to an endpoint of the deformable chain.  Without loss of generality, we will assume that, in the reference configuration, the leftmost atom on the deformable chain $ \mathcal C_2^0$ lies directly above an atom on $\mathcal C_1$.


%

\begin{figure}[htb]
\centering
\includegraphics[width=1.0\linewidth, clip, trim=1.2in .8in 0in 1in]
{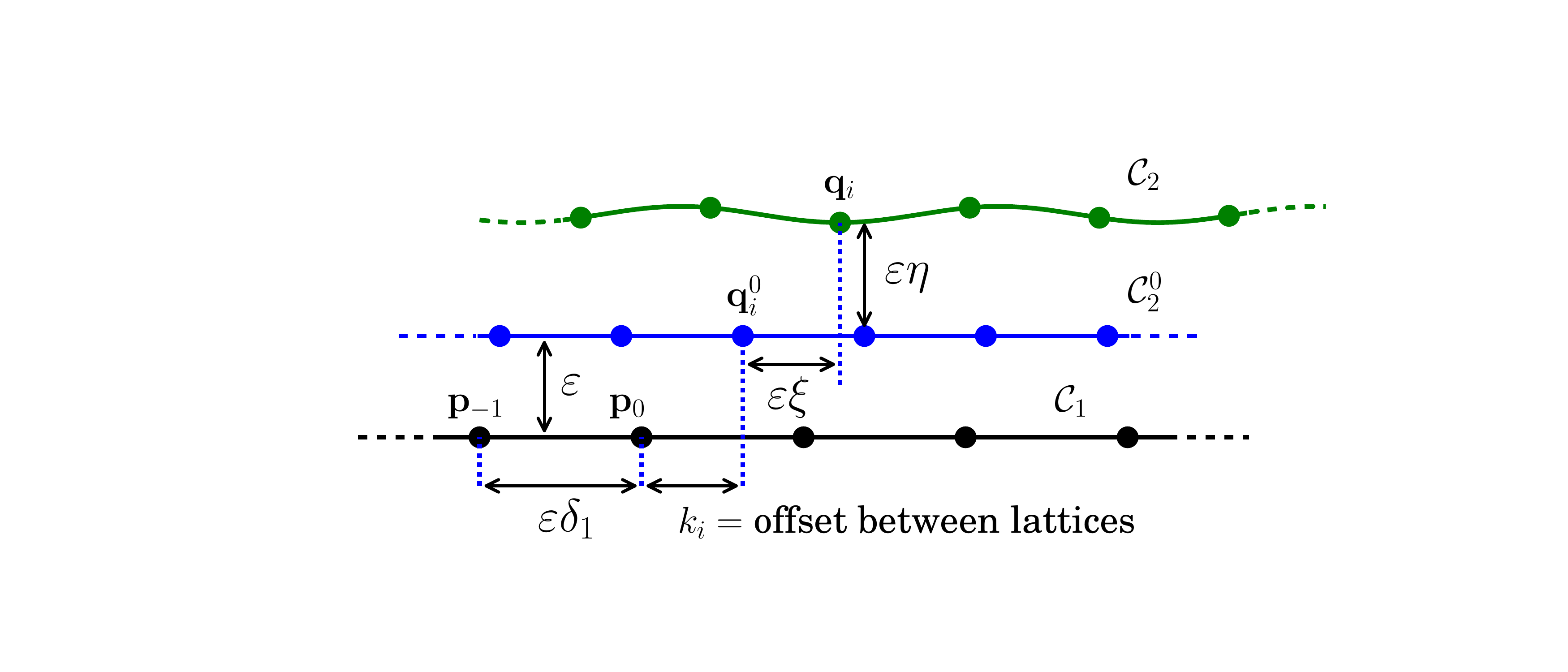}
\caption{Finding $d_{ij}$.}
  \label{f8}
\end{figure}

\begin{figure}[htb]
\hspace*{.05\linewidth}
\includegraphics[width=0.8\linewidth, clip, trim=0in .8in 1in 1in]
{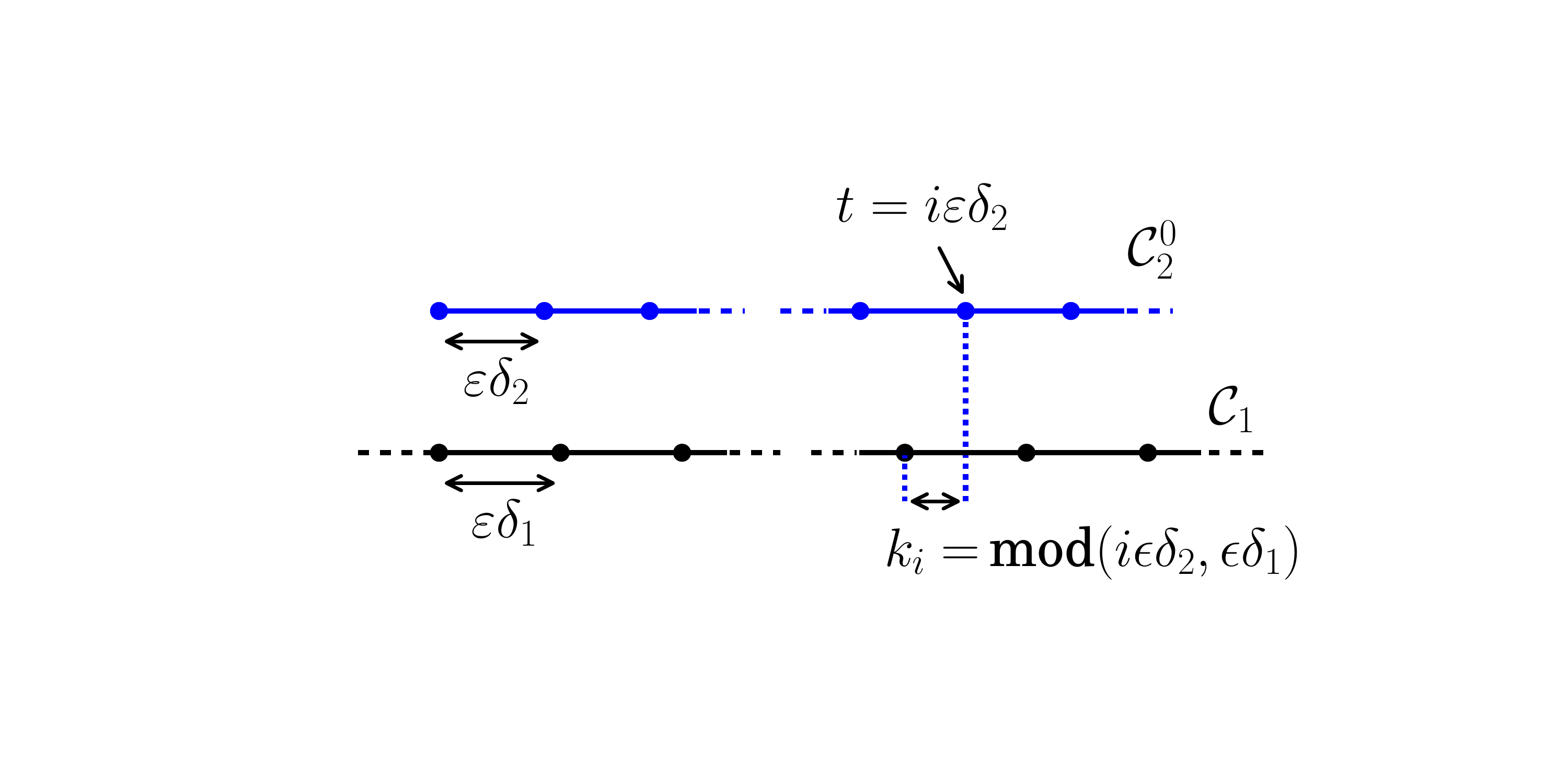}
\caption{Definition of the lattice offset.}
  \label{f10}
\end{figure}

To write down an expression for $d_{ij}$, we let $k_i$ denote the offset
between the atomic lattices of the two chains in the reference configuration as measured at the atom $i$ on
$\mathcal{C}_{2}^0$.  Then $k_i$ is the distance between the projection of
the atom $i$ onto $\mathcal{C}_{1}$ and the first atom on
$\mathcal{C}_{1}$ to the left of this projection.  We will label this atom on the rigid chain as the atom with the index $j=0$.
See Figure~\ref{f8}.  It is now clear that
\begin{equation}
  d_{ij}
  =
  \sqrt{\left(j\varepsilon\delta_{1}+k_i+\varepsilon\xi\left(\eps\delta_2i\right)\right)^{2}
  + \left(\varepsilon+\varepsilon\eta\left(\eps\delta_2i\right)\right)^{2}}.
  \label{ee15}
\end{equation}
Furthermore, since the atom at the left end of $\mathcal{C}_{2}^0$ is
directly above the $0$-th atom on $\mathcal{C}_{1}$, if the horizontal
component of the position
of the atom $i$ on $\mathcal{C}_{2}$ is $t_i= \varepsilon\delta_{2}
i$, then
\begin{equation}
  k_i
  =
  \text{mod}\,(i\varepsilon\delta_{2},\varepsilon\delta_{1})=i\varepsilon(\delta_{2}-\delta_{1})-i_1\varepsilon\delta_1
  = i\varepsilon\delta_{2}(\delta_{1}-\delta_{2})/\delta_{2}-i_1\varepsilon\delta_1,
  \label{ee4}
\end{equation}
for some $i_1\in\mathbb N$.
See Figure~\ref{f10}.
Hence for any $t$ we define 
\begin{equation}
   k(t) := \varepsilon\alpha t, \label{ee5}
\end{equation}
where $\alpha:=(\delta_{2}-\delta_{1})/\varepsilon\delta_{2}=\mathcal{O}(1)$. And therefore, our function $G$  that gives a
continuum description of the mismatch of the spacing between the atoms
on each chain is defined by
\begin{align}
	G(t,\xi(t),\eta(t))
        &=
        \sum_{j=-\infty}^{\infty}g(d_{ij}/\varepsilon) \nonumber \\
        &=
        \sum_{j=-\infty}^{\infty}
        g\left(
          \frac{\sqrt{((j-i_1)\,\varepsilon\delta_{1}+k(t)+\varepsilon\xi(t))^{2}
          + 
          (\varepsilon+\varepsilon\eta(t))^{2}}
          }{\varepsilon}\right) \label{ee8}\\
        &=
        \sum_{j=-\infty}^{\infty}
        g\left(
          \sqrt{(j\delta_{1}+\alpha t+\xi(t))^{2}
          + 
          (1+\eta(t))^{2}}
          \right), \nonumber 
\end{align}
where we translated the index $j$ by $i_1$ on the last step.

\subsection{Continuum Energy} \label{encon}

Putting together all the contributions, {one choice for} the continuum {energy} of the system is 
\begin{multline}
	\mathcal E^\eps[\xi,\eta] := 
        \frac{\gamma_{s}\varepsilon}{2}\int_{0}^{1}\!\left\{{\left(\xi^\prime(t)+\frac{\eps}{2}{\left(\eta^\prime(t)\right)}^2\right)}^2-\eps^2{\left(\xi^\prime(t)\eta^\prime(t)\right)}^2-\frac{\eps^2\delta_2^2}{12}{\left(\xi^{\prime\prime}(t)\right)}^2\right\}\,dt \\
        +
        \frac{\gamma_{b}\varepsilon^3}{2}\int_{0}^{1}\!\left(\eta''\right)^{2}\,dt
        +
        \frac{1}{\varepsilon}\!\int_{0}^{1}\!G(t,\xi,\eta)\,dt.
        \label{ee6.0}
\end{multline}

Note that the variational problem associated with the energy \eqref{ee6.0} is not well-posed since the negative sign in front of ${\left(\xi^{\prime\prime}\right)}^2$ leads to $\min{\mathcal E^\eps}=-\infty$. This issue is due to our choice to terminate the expansion \eqref{eq:bond} of the bond vector at $O\left(\eps^4\right)$; once the additional terms are included, the coefficient in front of the term containing the third derivative of $\xi$ squared will be positive. The downside of incorporating higher derivatives, however, is that the model quickly becomes more complicated. 

From the Ginzburg-Landau structure of the energy in \eqref{ee6.0}, if we expect the minimizers of \eqref{ee6.0} to develop wrinkles of characteristic width $\eps$, then we should expect all derivatives of the minimizers to appear as powers of $\eps^{-1}$ inside the wrinkled regions. Accordingly, all elastic terms in \eqref{ee6.0} will then contribute roughly the same amount to the overall energy. In fact, a quick glance at \eqref{eq:bond}, indicates that all terms in that expansion can also have the same magnitude inside the wrinkles, {possibly invalidating} the asymptotic procedure that led to \eqref{ee6.0}. 

On the other hand, in the regions between the wrinkles, all minimizers should have bounded derivatives and the terms with higher powers of $\eps$ in \eqref{ee6.0} should simply provide small corrections to the lower order contributions. The situation is not unlike that arising in continuum modeling of crystalline solids, where the structural defects---such as dislocations---can only be described in terms of their influence on the global strain field, without properly resolving the cores of the defects. {In order to fully resolve these cores, an alternative approach would be to use a quasicontinuum method that combines continuum description in the bulk with the discrete description in a vicinity of each defect \cite{blanc_lebris_lions_2007}.}

{With these observations in mind, we will adopt the following strategy to formulate a purely continuum theory.} Since the negative terms in \eqref{ee6.0} should only contribute inside the wrinkles and since the asymptotic procedure that led to \eqref{ee6.0} is likely not to be valid inside the wrinkled regions, we neglect these terms from now on. We conjecture that the overall effect of this change on the structure of minimizers will be restricted to a perturbation in the cross-section of the wrinkles. {Numerical simulations in the next section appear to support this statement and, in fact, show that the predictions of the proposed continuum model are close to the results of discrete simulations, even inside the wrinkles.} We set
\begin{align}
	\mathcal E^\eps[\xi&,\eta] := 
        \frac{\gamma_{s}\varepsilon}{2}\int_{0}^{1}\!\left(\xi'+\frac{\varepsilon}{2}(\eta')^{2}\right)^{2}
        \,dt
        +
        \frac{\gamma_{b}\varepsilon^3}{2}\int_{0}^{1}\!\left(\eta''\right)^{2}\,dt
        +
        \frac{1}{\varepsilon}\!\int_{0}^{1}\!G(t,\xi,\eta)\,dt.
        \label{ee6}
\end{align}
The energy functional in \eqref{ee6} can be viewed as a
generalization of similar energies for one-dimensional
Frenkel-Kontorova chains with the elastic contribution similar to that in F\"oppl-von
K\'arm\'an theory.

{Note that both the discrete and continuum nondimensional models contain the small parameter $\varepsilon$ and, in particular, the continuum model cannot be thought of as a limit of the discrete model as $\varepsilon\to0$. Instead, we conjecture that both models converge to the same asymptotic limit as $\varepsilon\to0$ in the appropriate sense. The limit has to be understood within the framework of $\Gamma$-convergence \cite{braides2002gamma} so that both energies are $\Gamma$-{\it equivalent} \cite{braides2008asymptotic}. Hence the number of terms retained in the expansion of the discrete problem in order to obtain the continuum problem should be sufficient to reproduce the behavior of the discrete system for a small $\varepsilon$. The proof of $\Gamma$-equivalence is a subject of a future work.}


The Euler-Lagrange equations for \eqref{ee6} are
\begin{align}
	-\gamma_{s}\varepsilon (\xi'' + \varepsilon \eta'\eta'')
        +
        \varepsilon^{-1}G_{\xi}(t,\xi,\eta)
        &=0, \label{e16}\\
	-\gamma_{s}\varepsilon^2\left[\left(\xi' + \frac{\varepsilon}{2}(\eta')^{2}\right)\eta'\right]'
        +
        \gamma_{b}\varepsilon^3\eta''''
        +
        \varepsilon^{-1}G_{\eta}(t,\xi,\eta)
        &=0. \label{e17}        
\end{align}
{Here \eqref{e16} corresponds to the horizontal force balance and the function $G_\xi$ describes how the tensile/compressive force in the chain changes with axial position. Similarly, \eqref{e17} is the vertical force balance, including the vertical force $G_\eta$ that arises from coupling between the chains mismatch and the out-of-plane displacement of the deformable chain. Note that the vertical deformation in the von K\'arm\'an contribution to \eqref{ee6} is multiplied by the small parameter $\eps$, in contrast with a typical form of this expression. } 

The natural boundary conditions are as follows
\begin{gather}
  \xi'(0)+\frac{\varepsilon}{2}(\eta'(0))^{2}
  =
  \xi'(1)+\frac{\varepsilon}{2}(\eta'(1))^{2}=0,
  \label{e18}\\
  \eta''(0)=\eta''(1)=0,\label{e19}\\
  \eta'''(0)=\eta'''(1)=0.\label{e20}
\end{gather}
{These boundary conditions have natural interpretations in elasticity theory: \eqref{e18} corresponds to the assumption of no applied compression, while \eqref{e19} and \eqref{e20} indicate that there is no applied moment or shear force. The boundary conditions and the force balances \eqref{e16}-\eqref{e17} can be modified in the standard way to include applied loads and body forces.}

\section{Numerical Results} \label{s3}


In this section, we study the predictions of our continuum model by
numerically solving the 2-point boundary-value problem defined by the
Euler-Lagrange equations \eqref{e16} and \eqref{e17} with the boundary
conditions \eqref{e18}--\eqref{e20}.

We make several basic comparisons between the predictions of our
continuum model and {those} of the discrete model.  The discrete simulations are conducted by using
dissipation-dominated (gradient-flow) dynamics based on the discrete
energy described in Section~\ref{s5}.  Recall that this energy has 3
terms that correspond to the stretching \eqref{eq:DiscreteE_s} and
bending \eqref{e21} energies of the deformable chain and the van der
Waals interaction between the deformable and the rigid chains
\eqref{eq:DiscreteE_w}.  Assuming gradient-flow dynamics gives a
system of ordinary differential equations.  These are solved
numerically until the solution equilibrates, which yields solutions
that are minimizers of the discrete energy.
{As an initial condition for both the discrete and continuum simulations, we assume that the rigid and deformable chains of atoms are parallel and at a distance $\sigma$ apart.}

\begin{figure}[htb]
\centering
\includegraphics[width=\linewidth]
{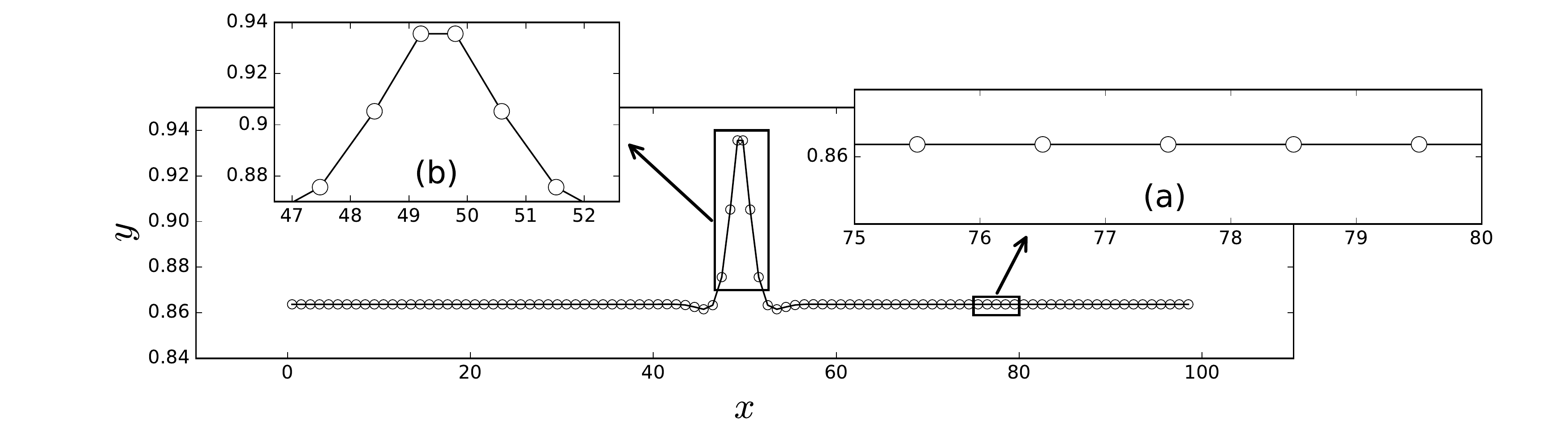} \\\hspace{-11mm}
\includegraphics[width=0.59\linewidth]
{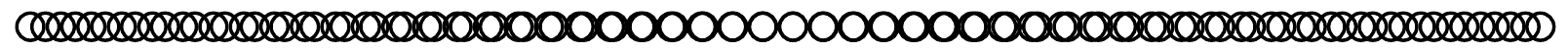}
\caption{Discrete simulation result with {$\sigma=1.0$, $\omega=1.0$, $h_{1}=1.0$, $h_{2}=.99$, $k_b=100.0$, and $k_s=10.0$}.  The atoms on the rigid chain, not shown, occupy positions
  $(j,0)$, where $j$ is an integer.  Inset (a) shows that in the
  commensurate regions, each atom falls into a potential well created
  by the rigid chain.  Inset (b) shows how the wrinkle allows six atoms to
  span over five potential wells. {The two lines of overlapping circles below the plot form the one-dimensional moir\'e pattern.}}
  \label{f20}
\end{figure}

Figure~\ref{f20} shows the result of an atomistic simulation for which
{$\sigma=1.0$, $\omega=1.0$, $h_{1}=1.0$, $h_{2}=.99$, $k_b=100.0$, and $k_s=10.0$}.  There are two relatively large
regions on which the atoms on the deformable chain are uniformly spaced and the
chain is parallel to the rigid chain.  These two flat regions are
separated by a single narrow region with a relatively large
vertical displacement, or wrinkle.  The atoms on the rigid chain are
at positions $(j,0)$ with $j$ an integer.  Hence the van der Waals
interaction from the rigid chain creates potential wells located above
every point $j+1/2$ on the $x$-axis.

As seen in inset (a) in Figure~\ref{f20}, in the regions where the
deformable chain is parallel to the rigid chain, the atoms fall into the potential
wells of the van der Waals interaction.  Hence in these regions the
atoms are 1 unit apart.  However, because $h_{2}=.99$, over the length
of the system there is one `extra' atom on the chain that has no
potential well.  As seen in inset (b) in Figure~\ref{f20}, this extra
atom is accommodated by a wrinkle, which minimizes the cost of placing
$n$ atoms over $n-1$ potential wells.  This wrinkle forms the domain
wall between the large commensurate regions to the left and the right.

{One way to understand the configuration of wrinkles that forms in a system of two incommensurate chains is to consider the corresponding moir\'e pattern. As noted in the introduction, when two lattices with different geometries or the same geometry but different orientations are stacked, a larger periodic pattern, called a moir\'e
pattern, emerges \cite{van2015relaxation,van2014moire}. A similar,
one-dimensional pattern arises when two incommensurate periodic chains
of atoms are overlaid. In Figure \ref{f20} this pattern is shown below
the plot of the discrete solution. The atoms of the rigid and deformable chains in the reference configuration are both represented by circles. At the left endpoint, the atoms of the deformable chain lie exactly above the midpoint between the atoms of the rigid chain, so that the corresponding circles show the minimum overlap. As one moves to the right along the chains, because of the difference between the lattice parameters $h_1$ and $h_2$, the atoms of the deformable chain progressively shift in relation to the atoms of the rigid chain until the circles lie exactly on the top of each other. This creates a pattern of lighter and darker regions, where the darker regions correspond to optimal atomic registry. On the other hand, the atoms of the deformable chain are out of registry in the lighter regions, and this is where the wrinkles form. Note that each light region corresponds to a single extra atom on the deformable chain, hence the number and locations of light regions in the moir\'e pattern should predict the number and the locations of wrinkles.  For instance, in Figure \ref{f20}, there is exactly one extra atom on the deformable chain and, as a consequence, exactly one light region and one wrinkle.}

\begin{figure}[H]
\centering
\includegraphics[width=0.4\linewidth]
{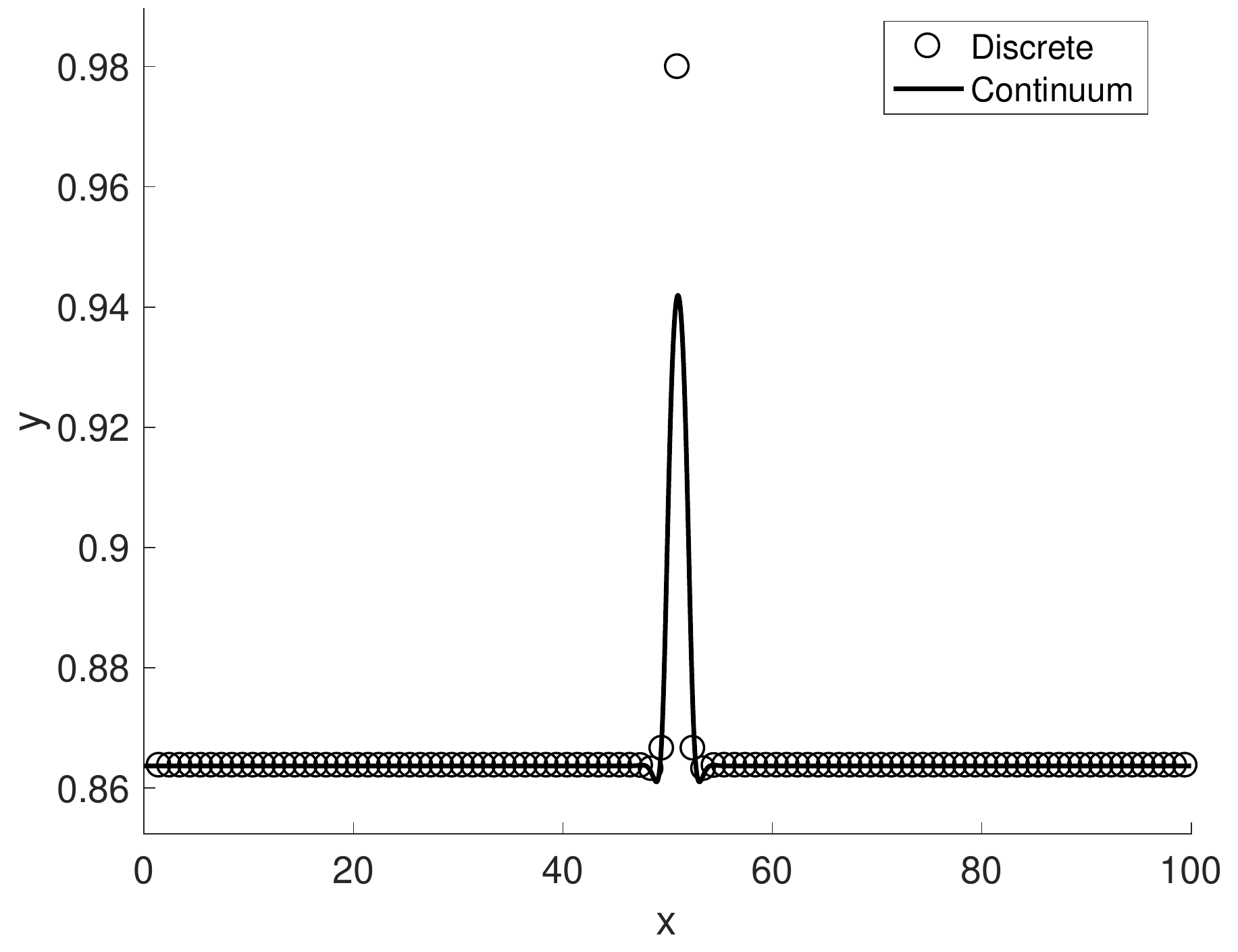}\qquad \includegraphics[width=0.39\linewidth]
{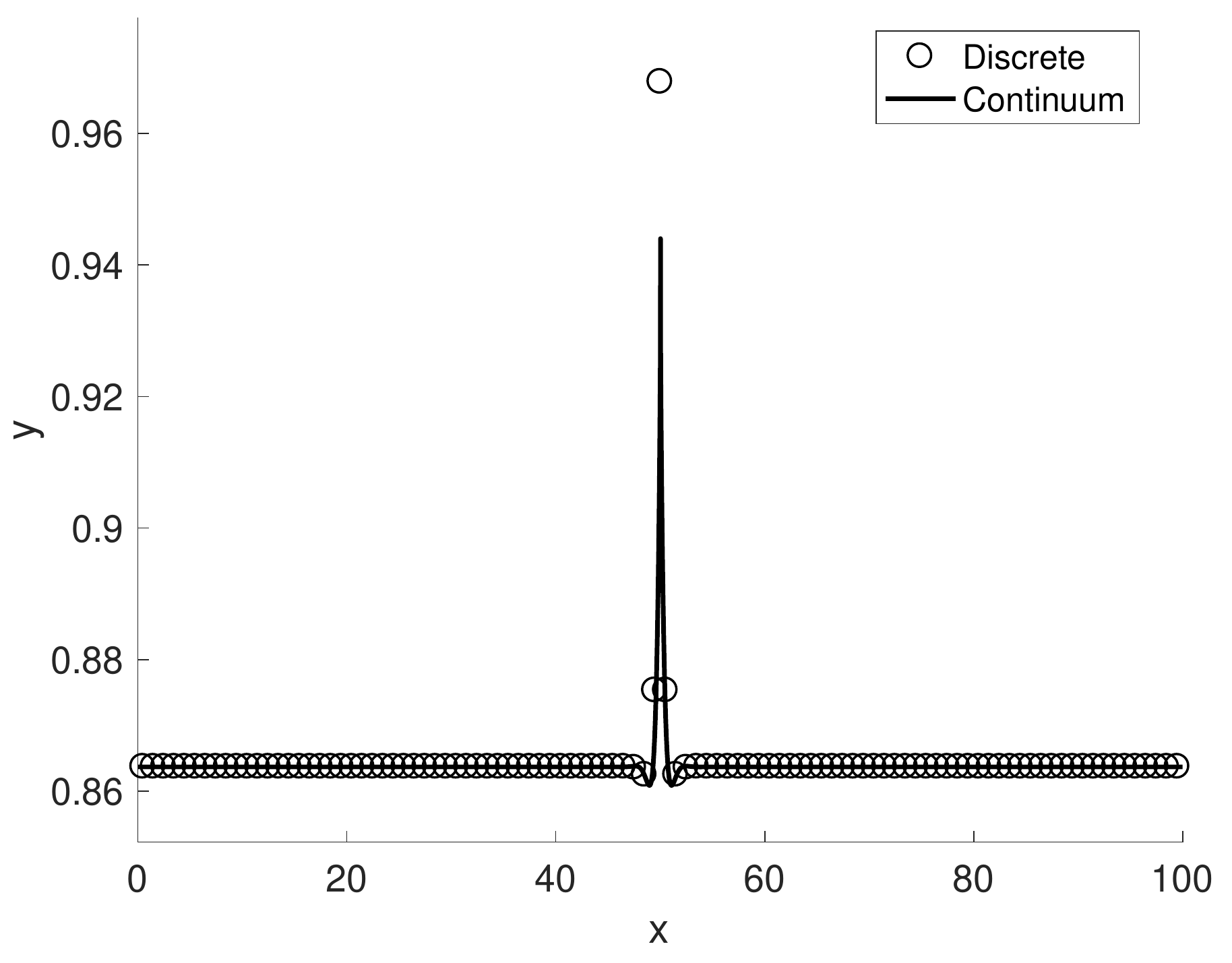}
\caption{{Discrete and continuum simulation results with $\sigma=1.0$, $\omega=1.0$, $h_{1}=1.0$, $k_s=1.0$ and $k_b=1.0$. The parameter $h_{2}=1.01$ and $h_{2}=.99$ on the left and right, respectively. The discrete simulation involves $100$ atoms on the rigid chain while there are $99$ (left) and $101$ (right) atoms on the deformable chain per period.}}
  \label{f23}
\end{figure}

\begin{figure}[H]
\centering
\includegraphics[width=0.4\linewidth]
{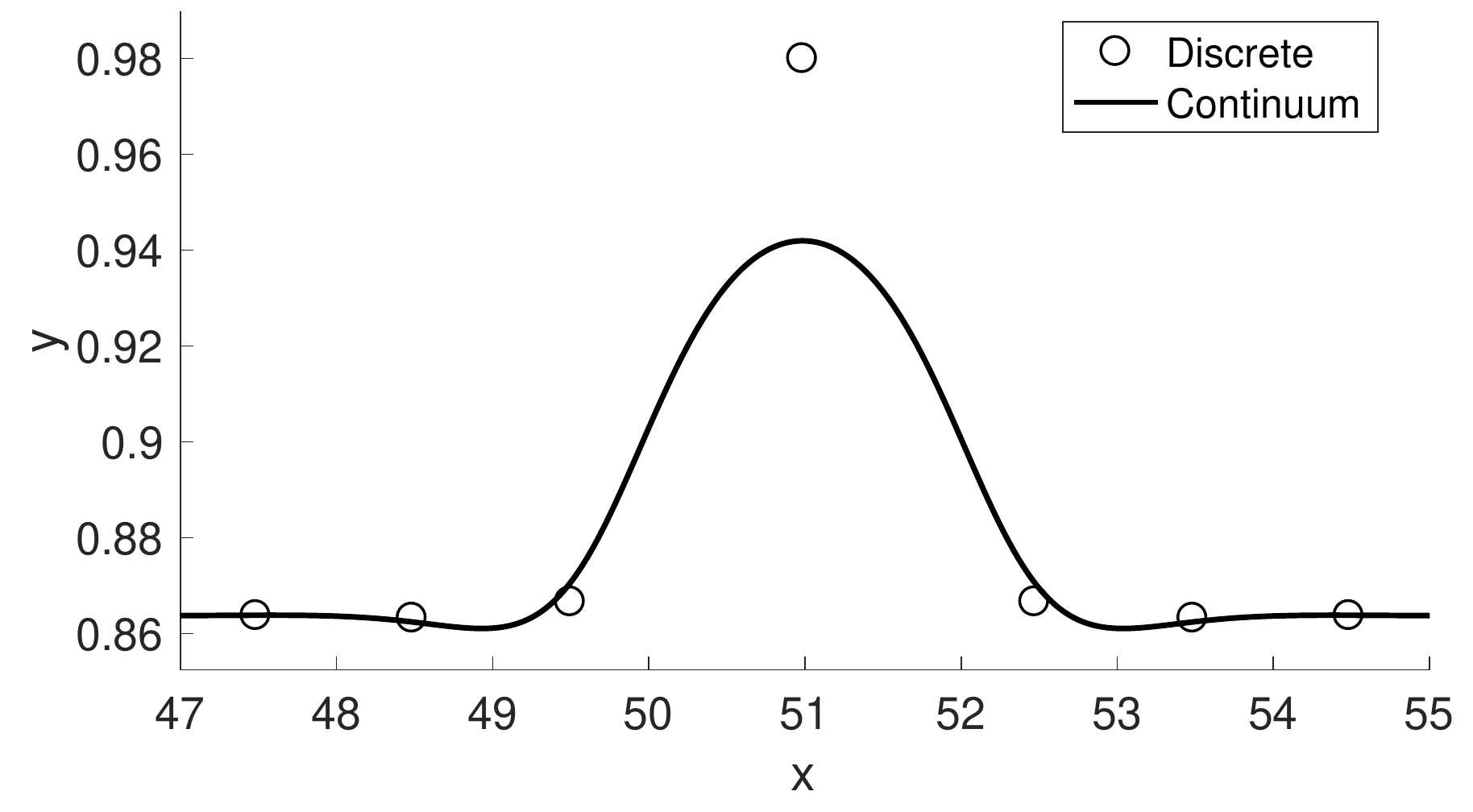}\qquad \includegraphics[width=0.39\linewidth]
{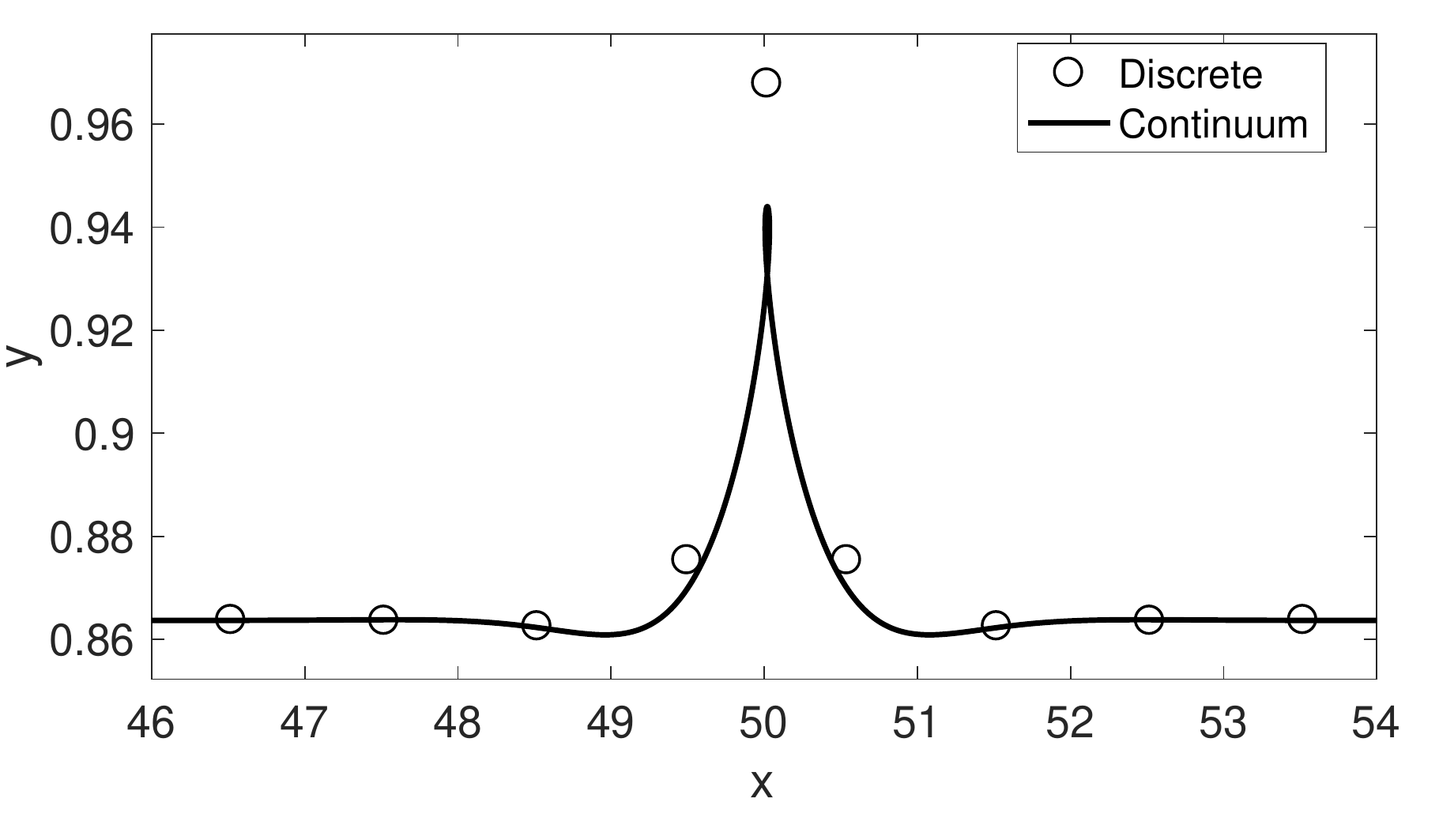}
\caption{{The details of atomistic structure near the wrinkle when $\sigma=1.0$, $\omega=1.0$, $h_{1}=1.0$, $k_s=1.0$ and $k_b=1.0$. The parameter $h_{2}=1.01$ and $h_{2}=.99$ on the left and right, respectively. The discrete simulation involves $100$ atoms on the rigid chain while there are $99$ (left) and $101$ (right) atoms on the deformable chain per period.}}
  \label{f23bis}
\end{figure}

{Figures~\ref{f23} and \ref{f23bis} show the numerical solutions of
both discrete and continuum models for parameter values 
\begin{equation}
\label{eq:comp}
\sigma=1.0,\ \omega=1.0,\ h_{1}=1.0,\ k_s=1.0,\ k_b=1.0
\end{equation}
and for two different values of $h_2$.  In both cases, we see that the continuum solution has the same structure as the discrete solution, with a single wrinkle separating two commensurate regions. The continuum solution is close to the discrete solution except that the amplitude of the wrinkle in the continuum solution is significantly smaller than that in the discrete solution.}

{To further understand the details within the wrinkle, we note that
when $h_1<h_2$, there are fewer atoms per period on the deformable
chain than there are atoms on the rigid chain. Consequently, the
deformable chain contracts between the wrinkles so that interatomic
distance in the deformable chain matches that in the rigid chain. The
contraction in the bulk causes the expansion of distances between
atoms of the deformable chain inside the wrinkles thus making the
wrinkles relatively wide (the left plot in Figure~\ref{f23bis}).
On the other hand, when $h_1>h_2$, there are more atoms per period on
the deformable chain, leading to bulk expansion
and wrinkles that are relatively narrow  (the right plot in Figure~\ref{f23bis}).}

\begin{figure}[htb]
\centering
\includegraphics[width=0.4\linewidth]
{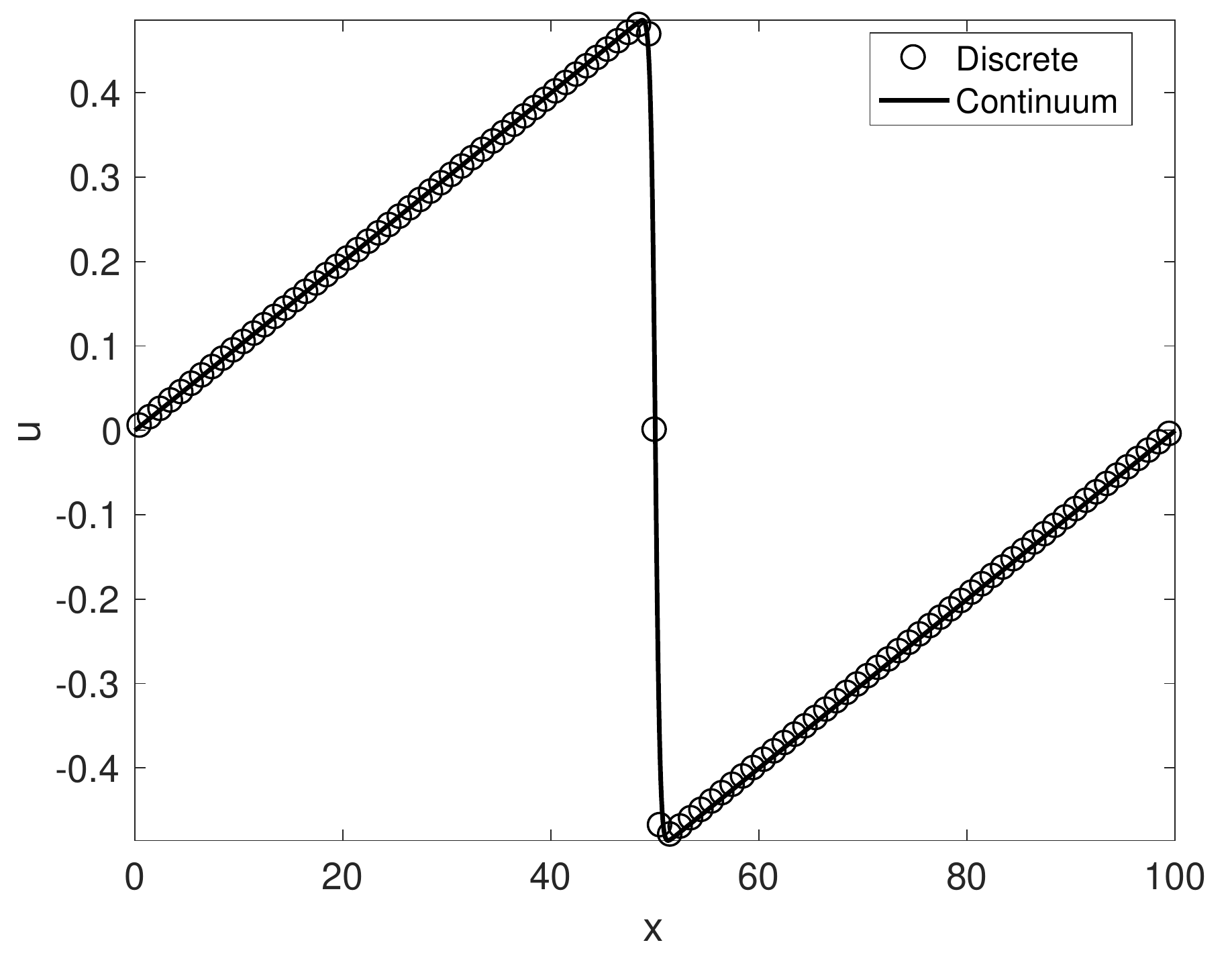}\qquad \includegraphics[width=0.41\linewidth]
{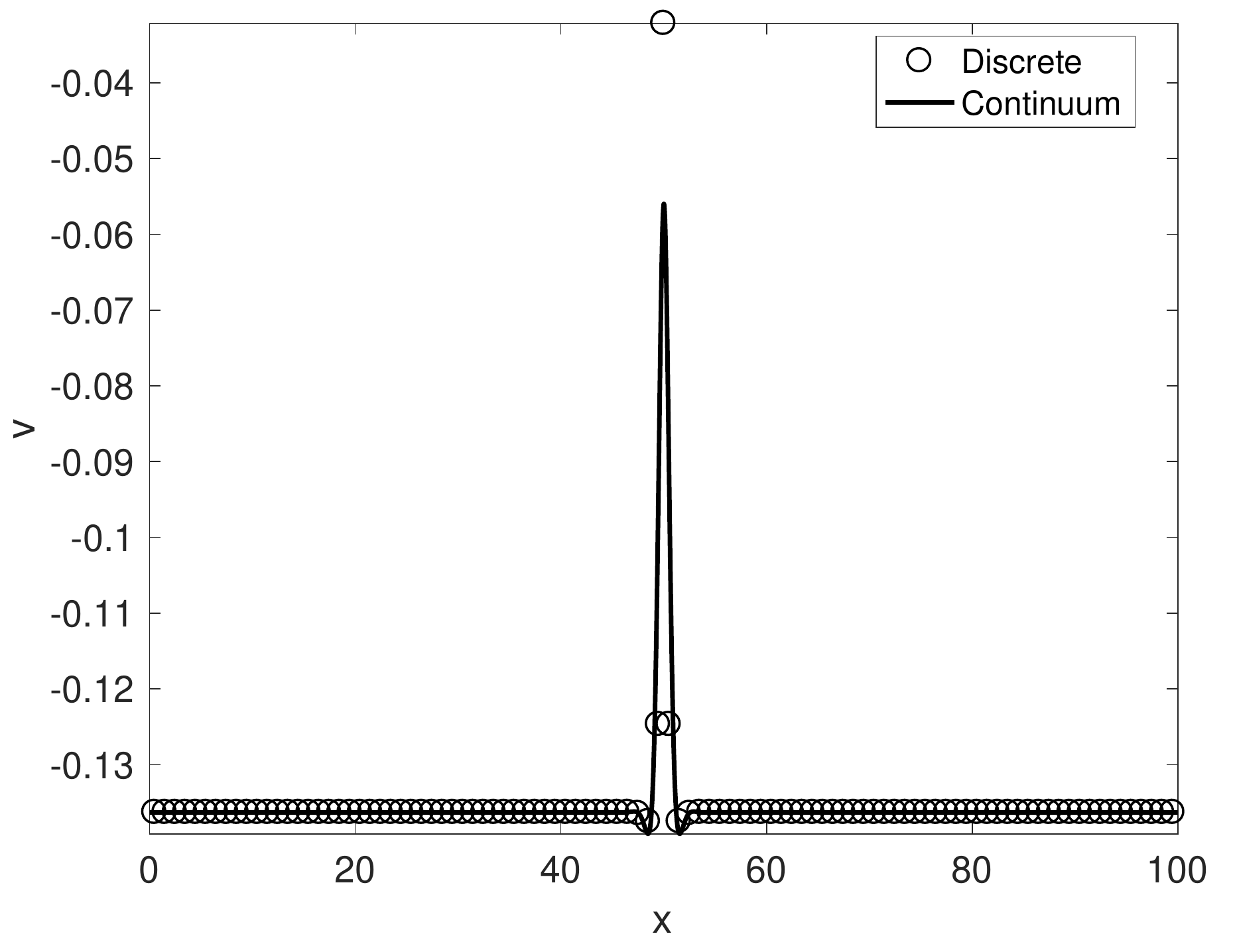}
\caption{{Discrete and continuum displacements for $\sigma=1.0$, $\omega=1.0$, $h_{1}=1.0$, $h_{2}=.99$, $k_s=1.0$ and $k_b=1.0$. The horizontal and vertical displacements are shown on the left and right, respectively. The discrete simulation involves $100$ atoms on the rigid chain while there are $101$ atoms on the deformable chain per period.}}
  \label{f23disp}
\end{figure}

{Note that only three atoms form the body of the wrinkle in the
  discrete solution in the two cases shown in Figures~\ref{f23} and \ref{f23bis}. This is to be expected since the nondimensional width
  of the wrinkle and the nondimensional interatomic distance are of
  order $\eps$ (these are of order $1$ in dimensional units). With so
  few atoms, there is no reason to expect that the continuum
  approximation should remain valid inside the wrinkles. The problem
  is further compounded by large gradients of the continuum solution
  in these singular regions. Indeed, as Figure~\ref{f23bis}
  demonstrates, the match between the discrete and continuum solutions
  is reasonable away from the center atom inside the wrinkle, while
  there is a significant discrepancy near the center atom. A further
  problem, associated with contraction inside the wrinkle, can be seen
  in the plot on the right in Figure~\ref{f23bis}.  The continuum
  solution forms a narrow loop near the maximum, indicating
  self-penetration of the deformable chain. Hence, the continuum solution for the combination of parameters
  \eqref{eq:comp} is unphysical near the top of the wrinkle. Note,
  however, that the nonphysicality is restricted to the region where
  there is only a single atom of the discrete solution. Indeed, as
  Figure~\ref{f23disp} demonstrates, the continuum displacements
  predict the discrete displacements well at the atomic positions for
  the values \eqref{eq:comp} of the parameters used to produce the
  plot on the right in Figure~\ref{f23bis}. We emphasize that the
  accuracy of the continuum approximation is the same for both plots
  in Figure~\ref{f23bis}. There is no self-penetration in the plot
  on the left simply because near the wrinkle the chain experiences local expansion rather than contraction.} 

{The elastic constants $k_s$ and $k_b$ are the same in
  \eqref{eq:comp}, so that the cost of stretching and bending is
  comparable for the solutions depicted in Figures~\ref{f23} and
  \ref{f23bis}. That bending is relatively expensive tends to spread
  bending deformation, causing the dips that form on both sides
  of the wrinkles as well as depressing the amplitude of the
  wrinkles themselves. In particular, large bending stiffness causes
  the deformable sheet to ``undershoot" the equilibrium distance
  between the chains in registry at the edge of a wrinkle. A
  subsequent ``bounce-off" away from the substrate is the cause of the
  dips adjacent to the wrinkles. A macroscopic version of this effect
  is well known for elastic beams on a liquid
  \cite{PhysRevLett.107.044301}.  An analysis similar to the one presented in \cite{PhysRevLett.107.044301} can be performed here to study the shape of the dips. Indeed, the ability to analyze the system of ordinary differential equations to establish the properties of equilibrium solutions is one of the clear advantages of working with the continuum rather than the discrete model.}

{We now fix the strength of the Lennard-Jones interaction
  $\omega=1$ and investigate the influence of the elastic constants on
  the discrete and continuum solutions. Generically speaking, we
  expect that with larger elastic constants the solution should incur
  a larger penalty for elastic deformation and, consequently, should
  exhibit smaller gradients. This in turn should lead to wider
  wrinkles that involve more atoms per wrinkle. Both effects should
  have a positive influence on the accuracy and applicability of the
  continuum approximation near the wrinkles.} 

\begin{figure}[htb]
\centering
\includegraphics[width=0.4\linewidth]
{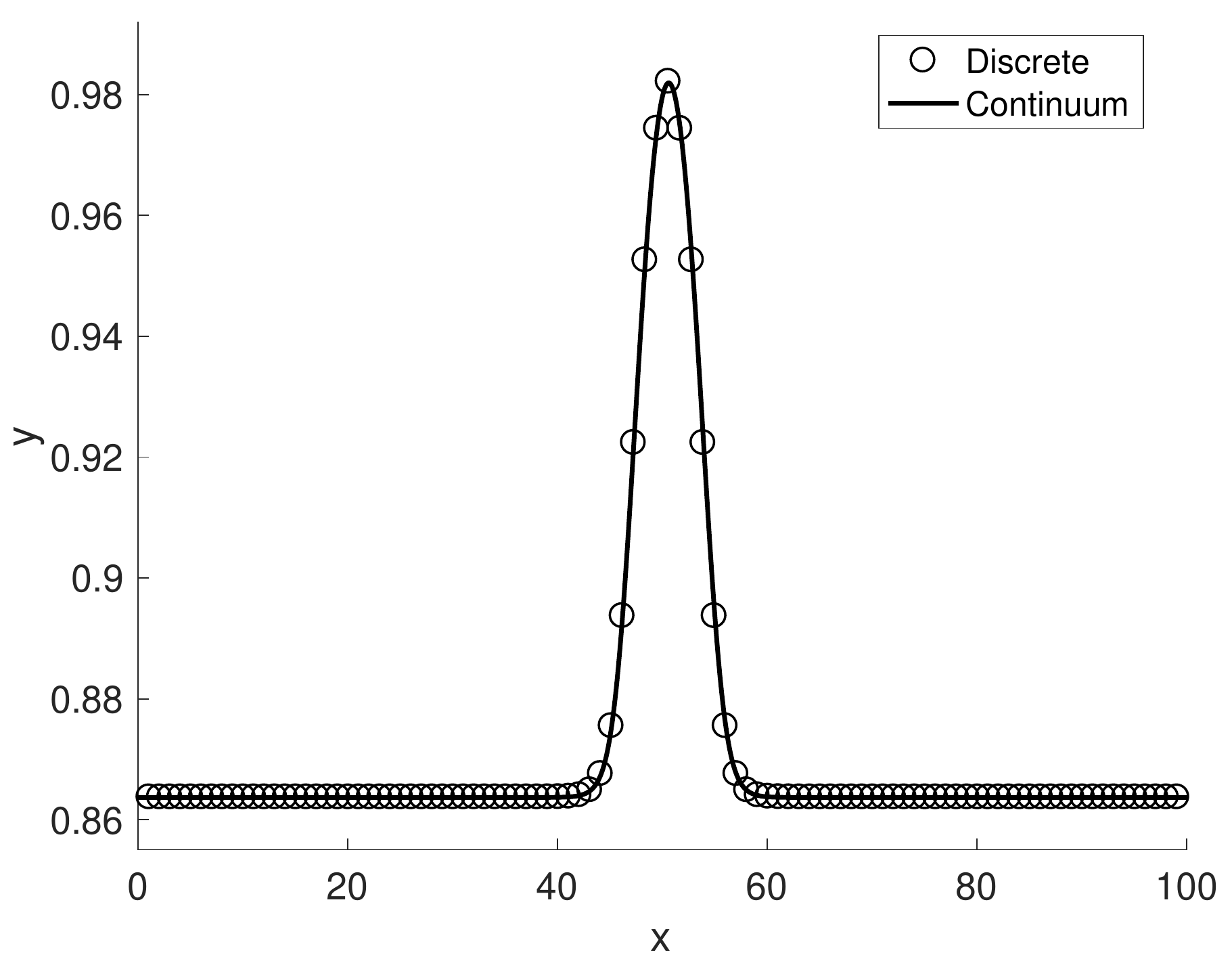}\qquad \includegraphics[width=0.39\linewidth]
{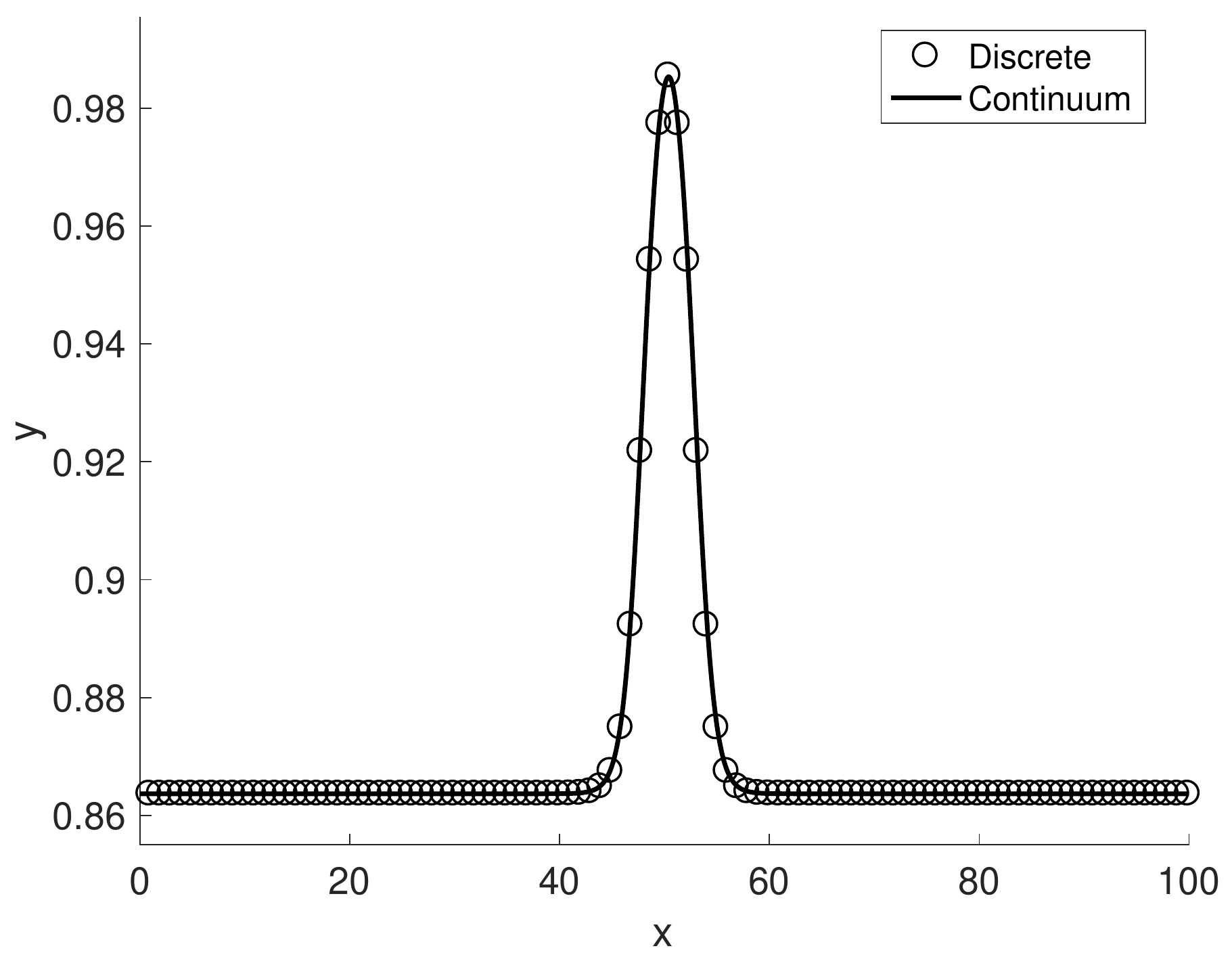}
\caption{{Discrete and continuum simulation results with $\sigma=1.0$, $\omega=1.0$, $h_{1}=1.0$, $k_s=100.0$ and $k_b=1.0$. The parameter $h_{2}=1.01$ and $h_{2}=.99$ on the left and right, respectively. The discrete simulation involves $100$ atoms on the rigid chain while there are $99$ (left) and $101$ (right) atoms on the deformable chain per period.}}
  \label{f23.1}
\end{figure}

{In Figure~\ref{f23.1}, we set $k_s=100.0$ and $k_b=1.0$ so that stretching deformation is significantly more expensive than bending. As a result, the bending stiffness is small and the dips next to the wrinkles disappear, both in the case of local stretching (Figure~\ref{f23.1}, left) and local compression (Figure~\ref{f23.1}, right). As expected, there are more atoms involved in forming a wrinkle, and the continuum approximation matches the discrete solution very well. Further, the amplitude of the wrinkles is larger than that observed in Figure~\ref{f23}. Note that, even though the number of atoms that form the wrinkle is now larger, the wrinkle itself still serves to accommodate only {\em one} extra atom on the deformable chain.}

\begin{figure}[htb]
\centering
\includegraphics[width=0.4\linewidth]
{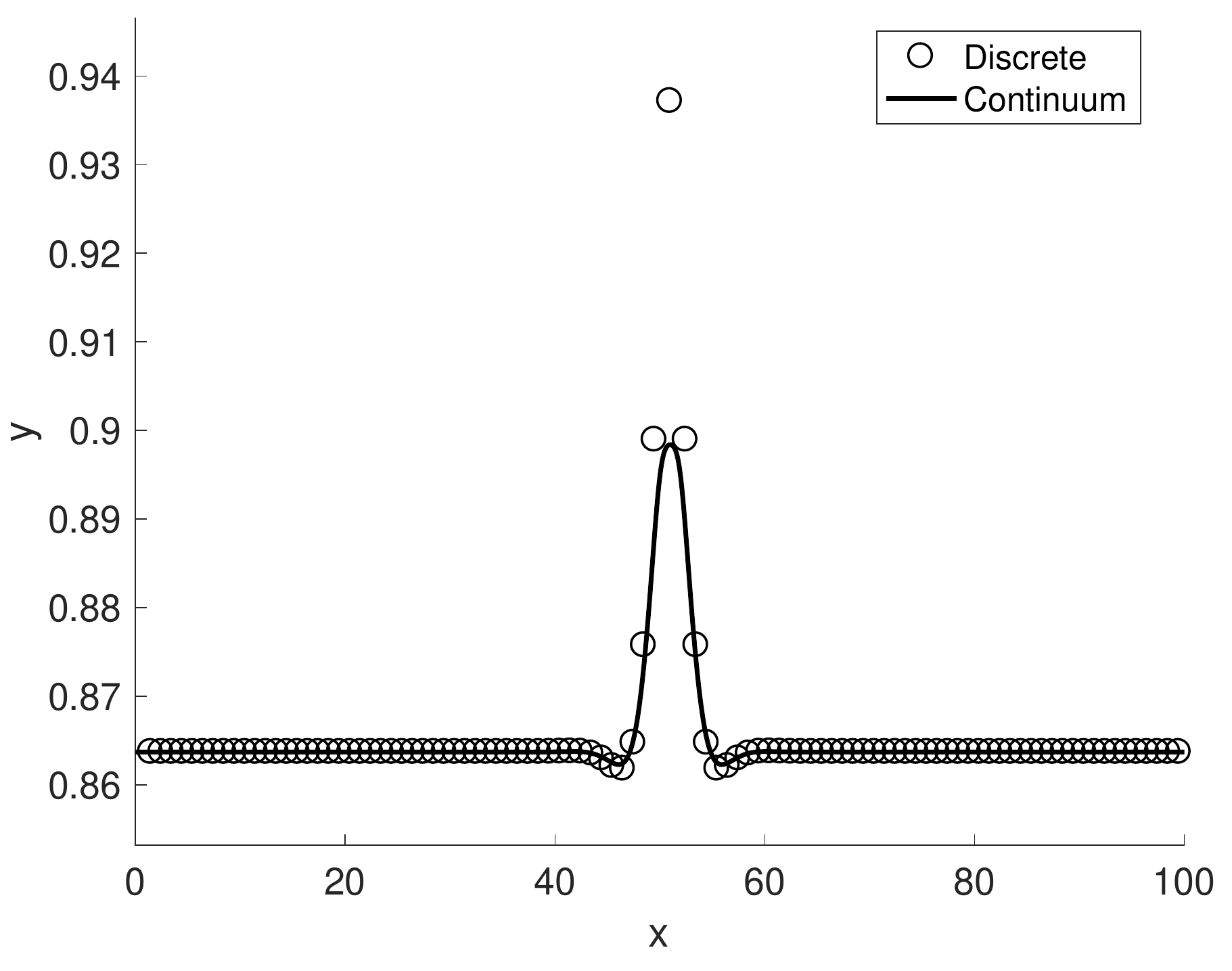}\qquad \includegraphics[width=0.39\linewidth]
{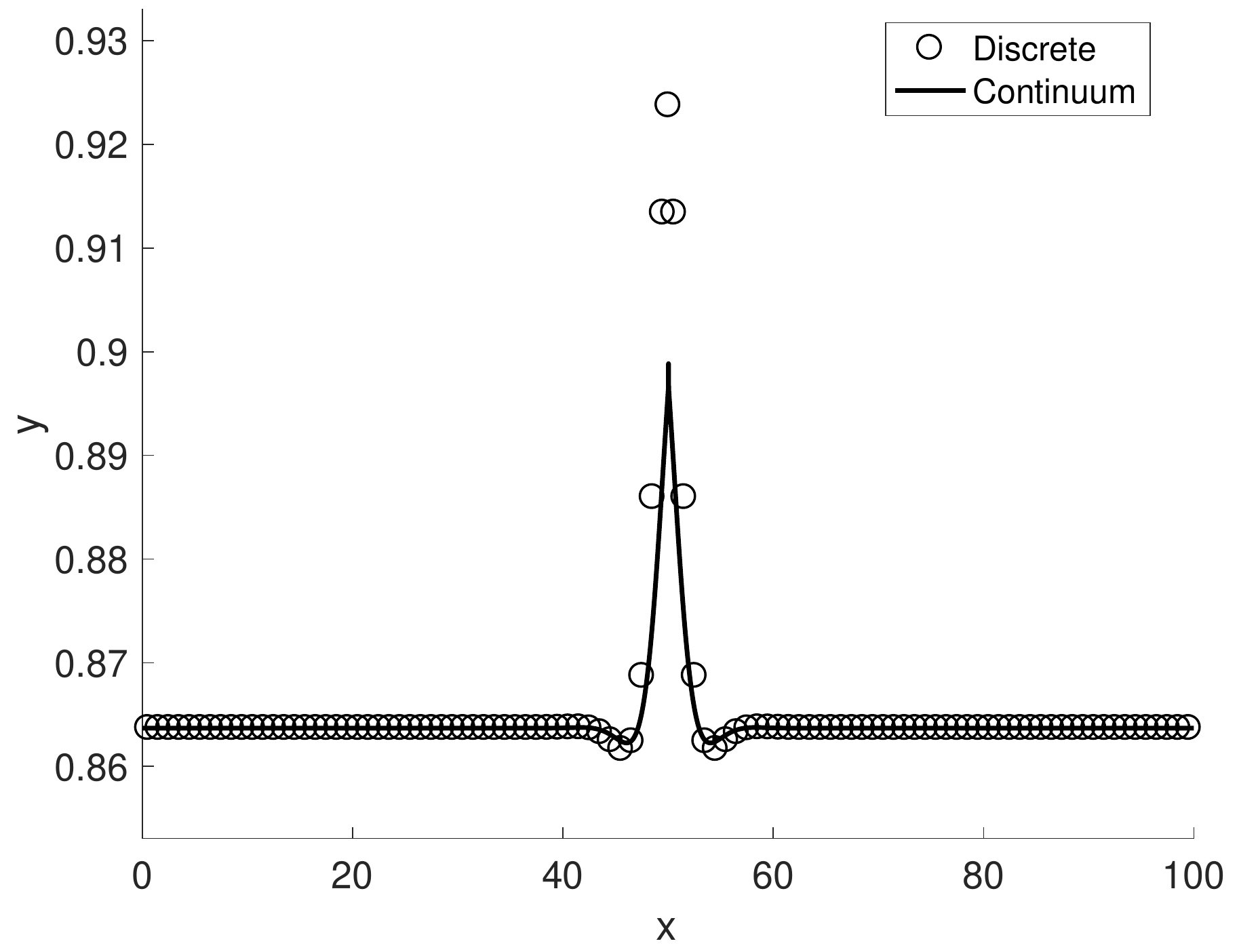}
\caption{{Discrete and continuum simulation results with $\sigma=1.0$, $\omega=1.0$, $h_{1}=1.0$, $k_s=1.0$ and $k_b=100.0$. The parameter $h_{2}=1.01$ and $h_{2}=.99$ on the left and right, respectively. The discrete simulation involves $100$ atoms on the rigid chain while there are $99$ (left) and $101$ (right) atoms on the deformable chain per period.}}
  \label{f23.1bis}
\end{figure}

{In Figure~\ref{f23.1bis}, we consider the opposite case, $k_s=1.0$
  and $k_b=100.0$, so that bending is significantly more expensive than
  stretching. Because the bending stiffness is now large, the dips
  next to the wrinkles are enhanced and the wrinkles have a smaller
  amplitude than in Figure~\ref{f23}. The combination of large bending stiffness
  and compression causes the self-penetration near the top of the
  wrinkle in the plot on the right in Figure~\ref{f23.1bis}.  Note that,
  typically, bending in two-dimensional materials is much cheaper than
  in-plane compression \cite{lu2009elastic,lee2008measurement}, so
  that in practice one is unlikely to encounter the extreme situation
  depicted in the plot on the right in Figure~\ref{f23.1bis}.}

\begin{figure}[htb]
\centering
\includegraphics[width=0.4\linewidth]
{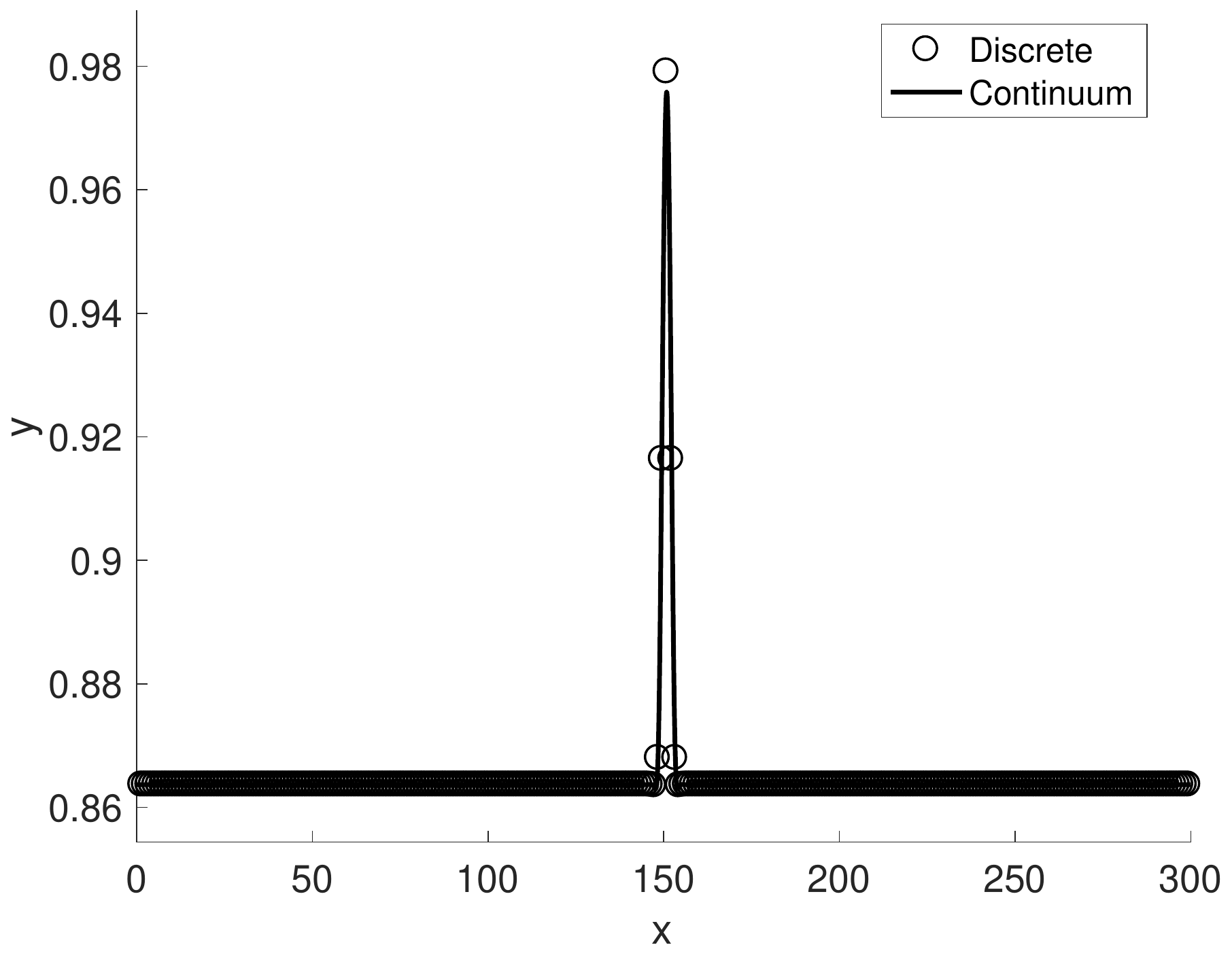}\qquad\includegraphics[width=0.4\linewidth]
{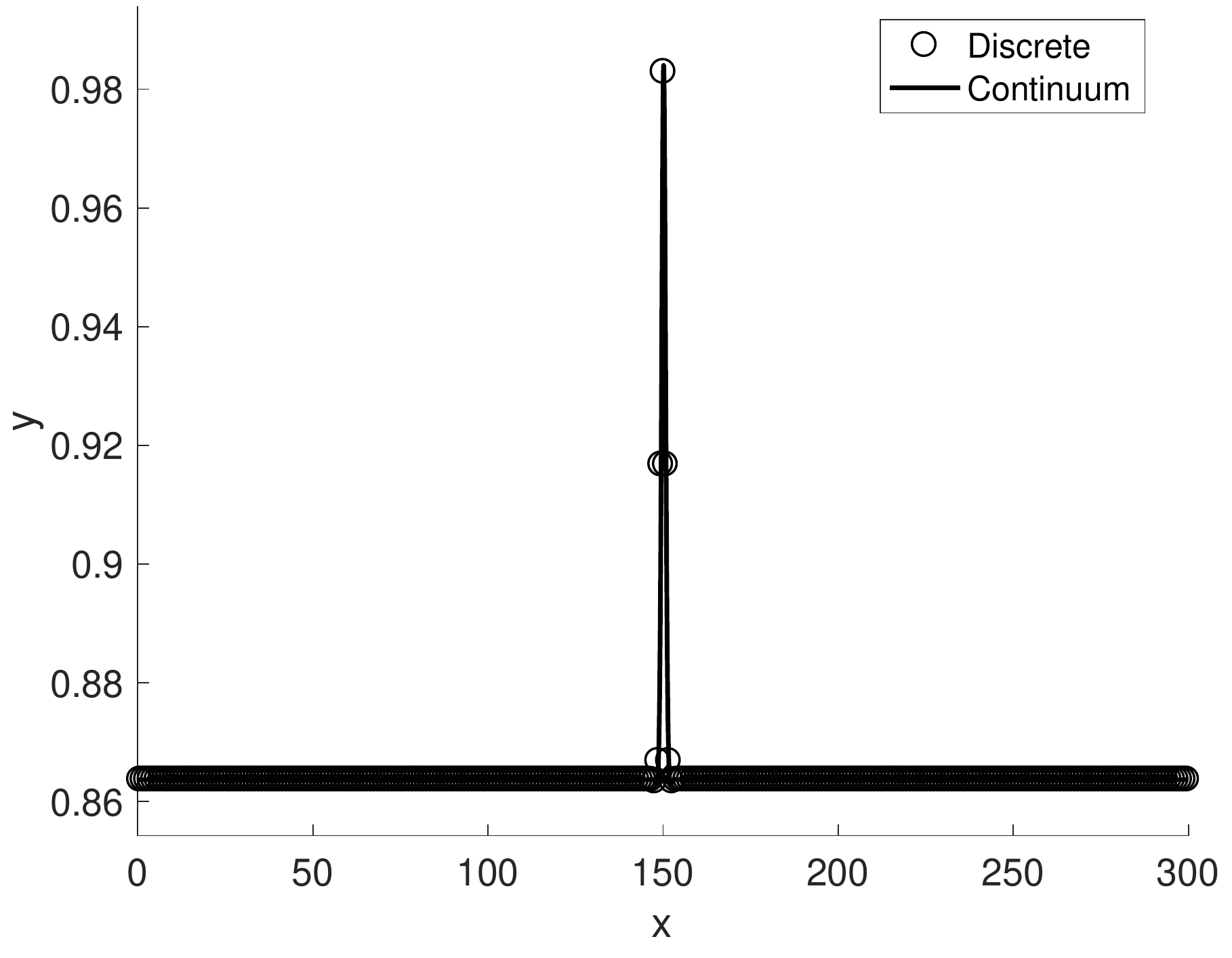}
\caption{{Discrete and continuum simulation results with $\sigma=1.0$, $\omega=1.0$, $h_{1}=1.0$, $k_s=10.0$ and $k_b=1.0$. The parameter $h_{2}=1.003$ and $h_{2}=.9967$ on the left and right, respectively. The discrete simulation involves $300$ atoms on the rigid chain while there are $299$ (left) and $301$ (right) atoms on the deformable chain per period.}}
  \label{f23.2}
\end{figure}

\begin{figure}[htb]
\centering
\includegraphics[width=0.4\linewidth]
{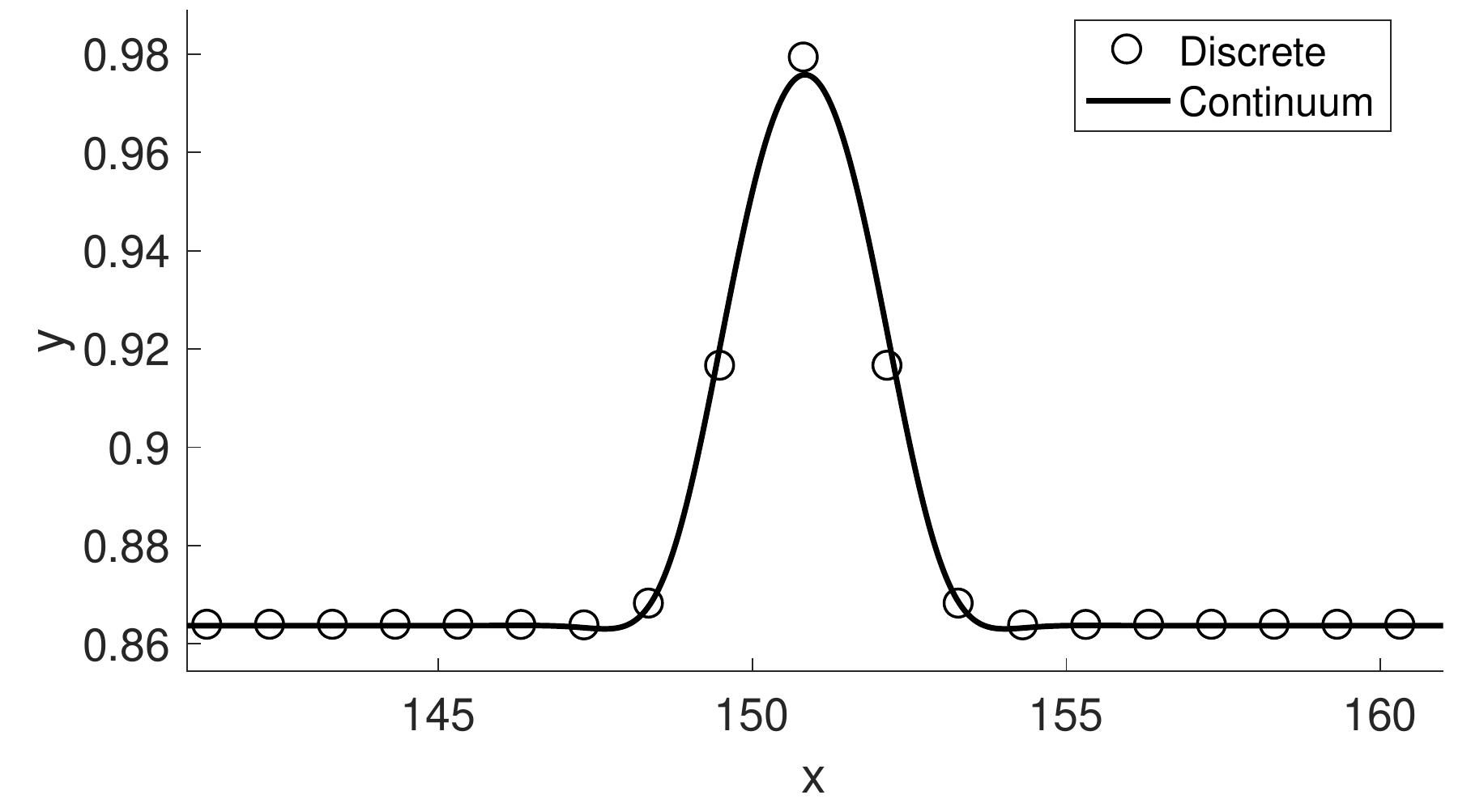}\qquad\includegraphics[width=0.4\linewidth]
{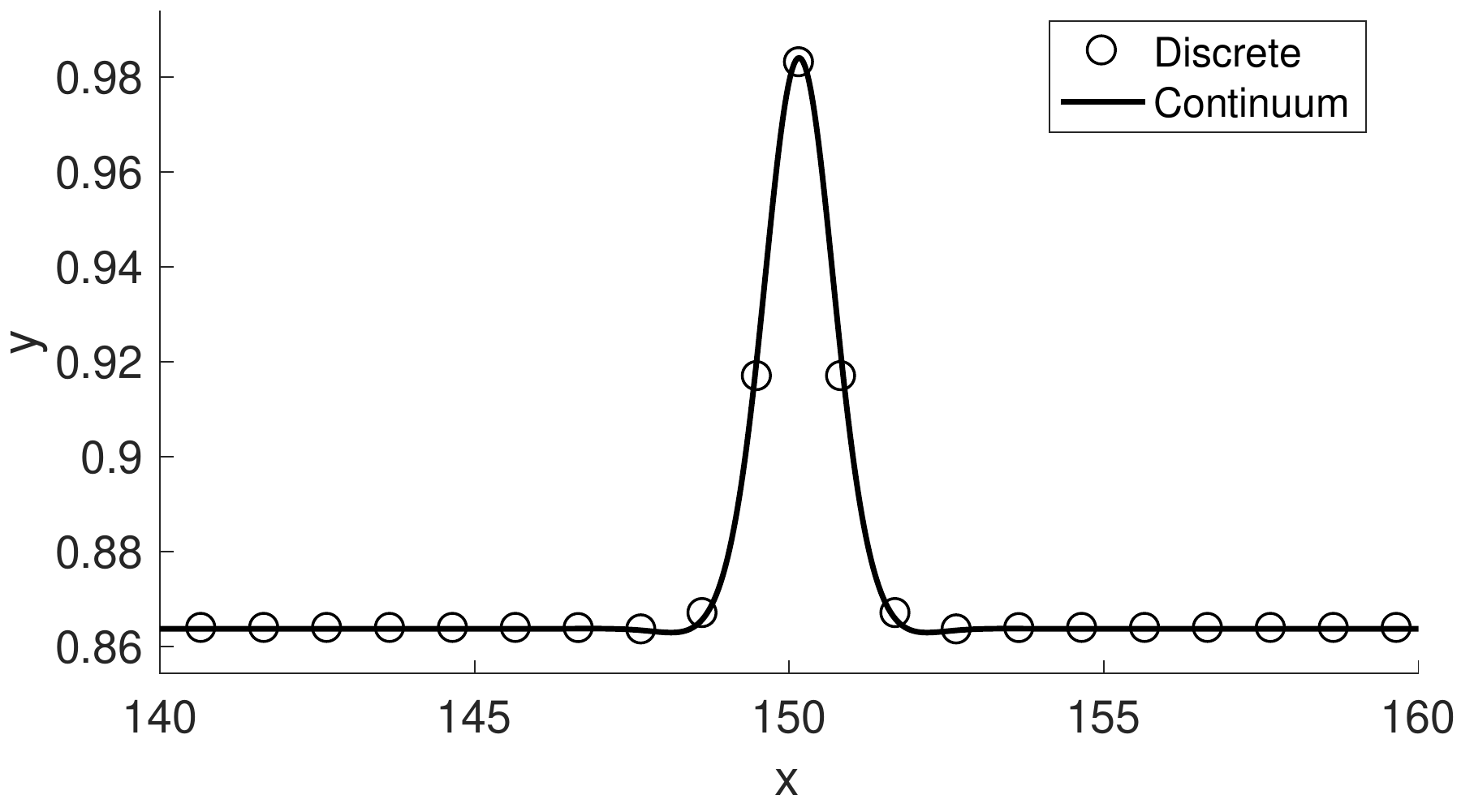}
\caption{{The details of atomistic structure near the wrinkle when $\sigma=1.0$, $\omega=1.0$, $h_{1}=1.0$, $k_s=10.0$ and $k_b=1.0$. The parameter $h_{2}=1.003$ and $h_{2}=.9967$ on the left and right, respectively. The discrete simulation involves $300$ atoms on the rigid chain while there are $299$ (left) and $301$ (right) atoms on the deformable chain per period.}}
  \label{f23.3}
\end{figure}

\begin{figure}[htb]
\centering
\includegraphics[width=0.4\linewidth]
{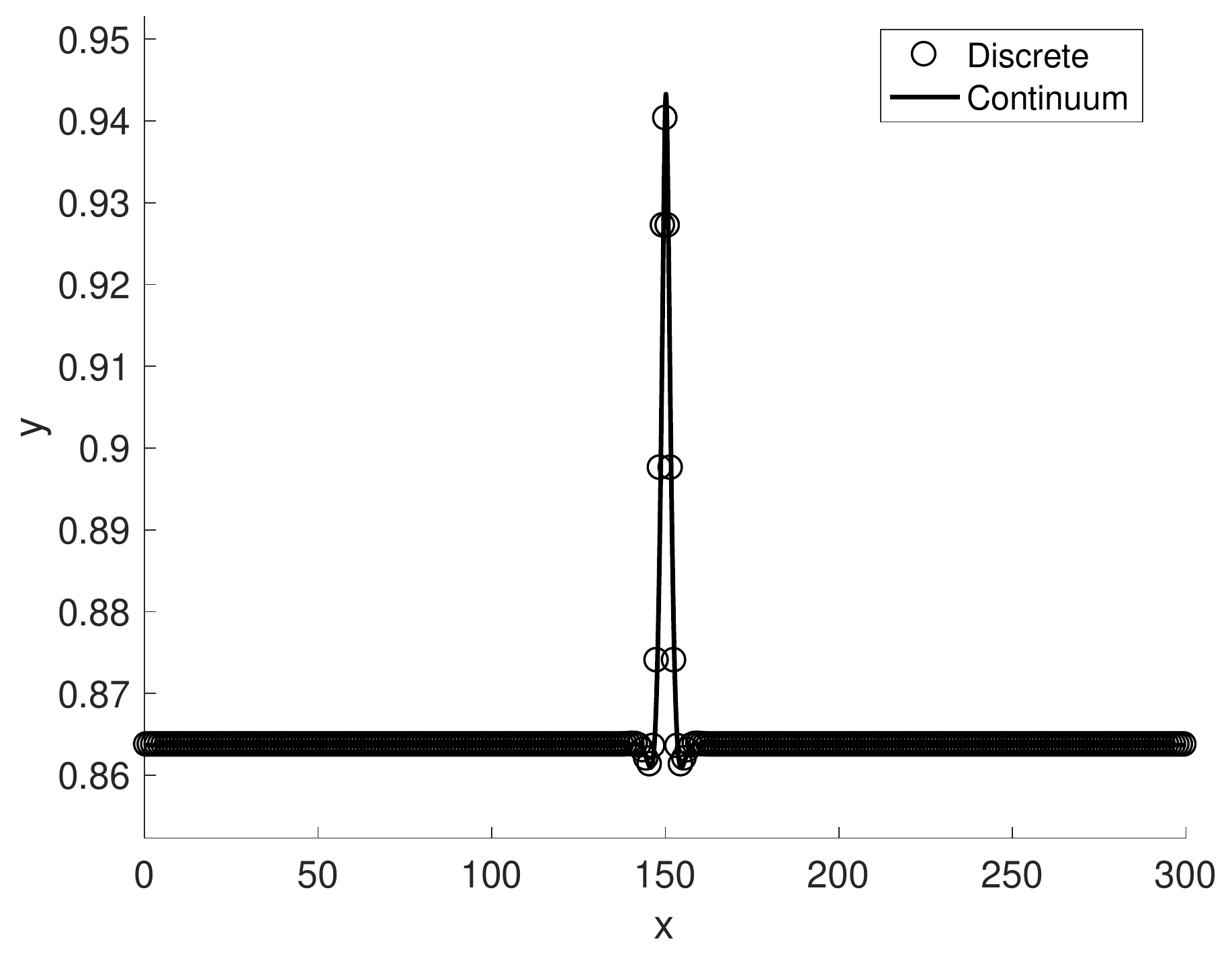}\qquad\includegraphics[width=0.55\linewidth]
{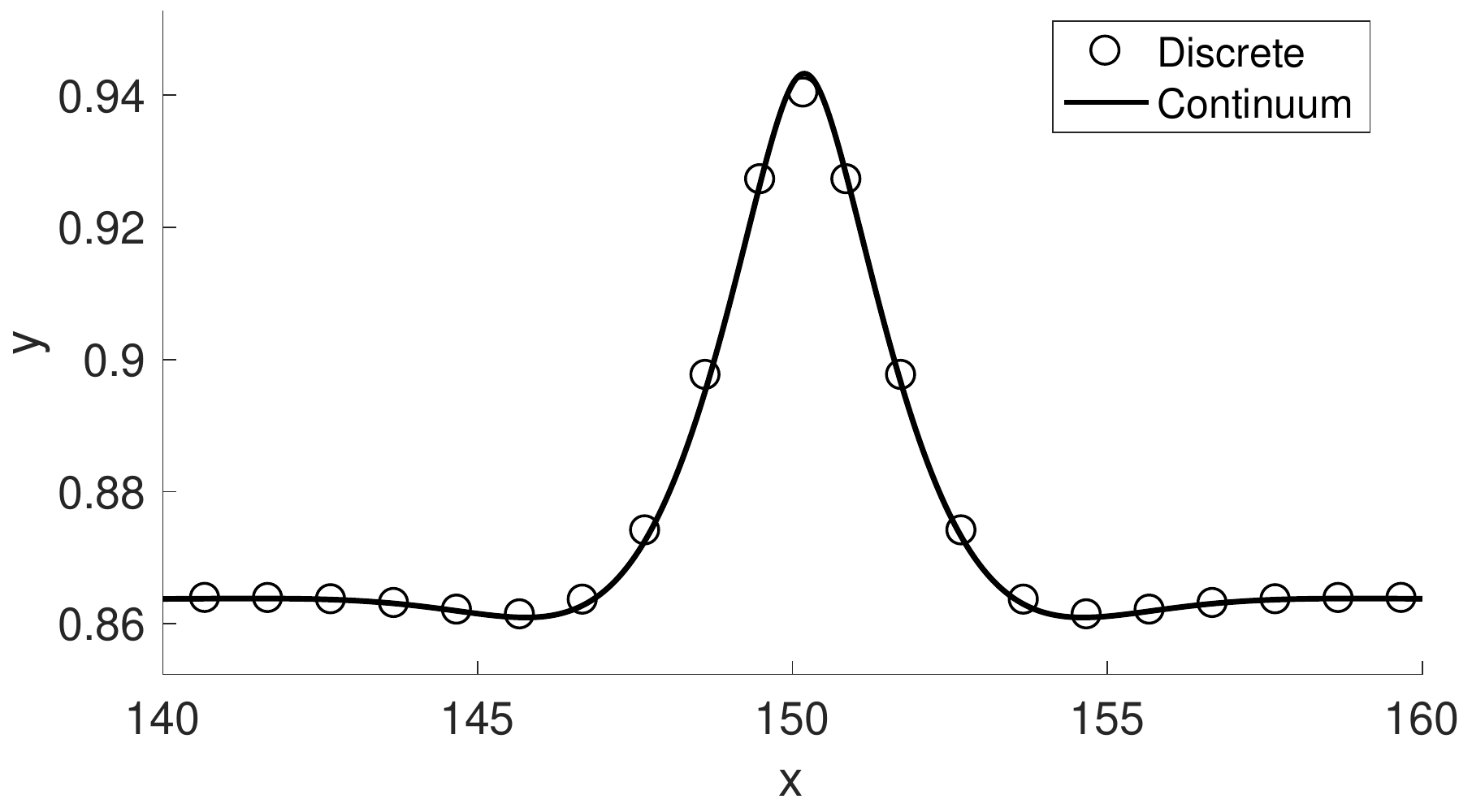}
\caption{{Discrete and continuum simulation results with $\sigma=1.0$, $\omega=1.0$, $h_{1}=1.0$, $h_{2}=.9967$, $k_s=10.0$, and $k_b=100.0$. The plot on the left shows the simulation results over a full period, while the plot on the right shows the details of atomistic structure near the wrinkle. The discrete simulation involves $300$ atoms on the rigid chain and $301$ atoms on the deformable chain per period.}}
  \label{f23.2bis}
\end{figure}

{The solutions for smaller value of the parameter $\eps$---when the size of the system is assumed to be larger given the same interatomic distance---are shown in Figures~\ref{f23.2}--\ref{f23.2bis}. For the relatively large stretching constant, $k_s=10.0$, it appears that all continuum solutions match well their discrete counterparts.}

\begin{figure}[h!]
\centering
\includegraphics[width=0.6\linewidth]
{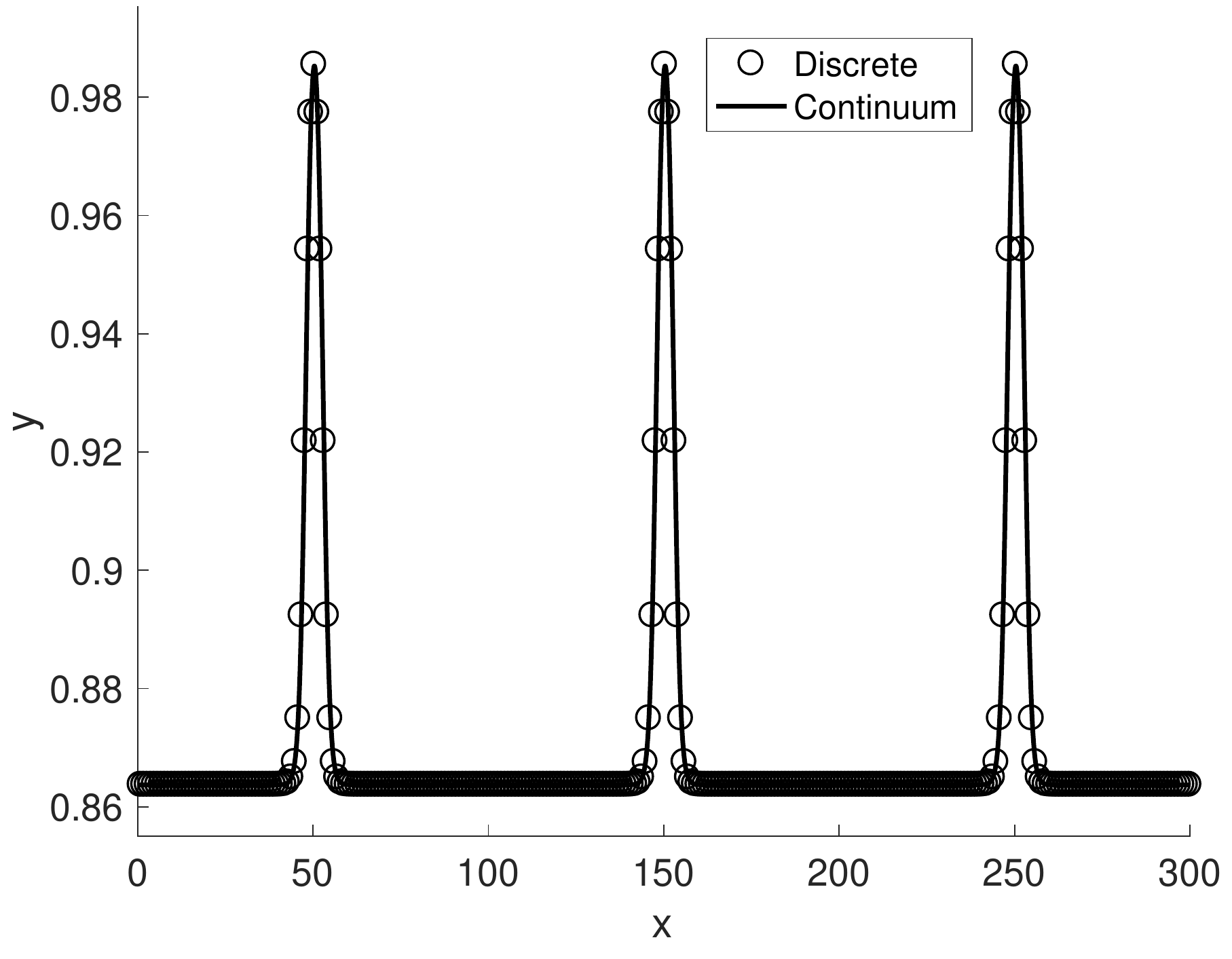}\\\hspace{7mm}
\includegraphics[width=0.54\linewidth]
{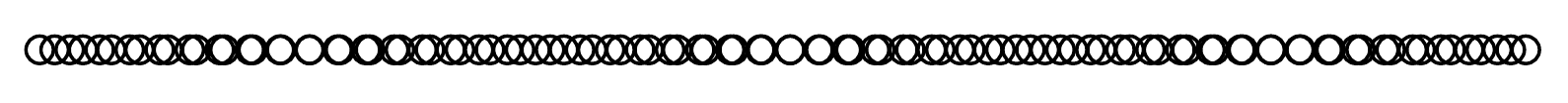}\vspace{1ex}
\caption{{Discrete and continuum simulation results with $\sigma=1.0$, $\omega=1.0$, $h_{1}=1.0$, $h_{2}=.99$, $k_s=100.0$, and $k_b=1.0$. The discrete simulation involves $300$ atoms on the rigid chain and $303$ atoms on the deformable chain per period. There are three wrinkles to accommodate the three `extra' atoms on the chain. The two lines of overlapping circles below the plot form the one-dimensional moir\'e pattern.}}
  \label{f29}
\end{figure}

{If the deformable chain has three extra atoms over the length of the system,
then three equispaced wrinkles form as shown in Figure \ref{f29}. The locations and the number of wrinkles are predicted by the moir\'e pattern at the bottom of Figure \ref{f29}.}

\begin{figure}[h!]
\centering
\includegraphics[width=0.5\linewidth]
{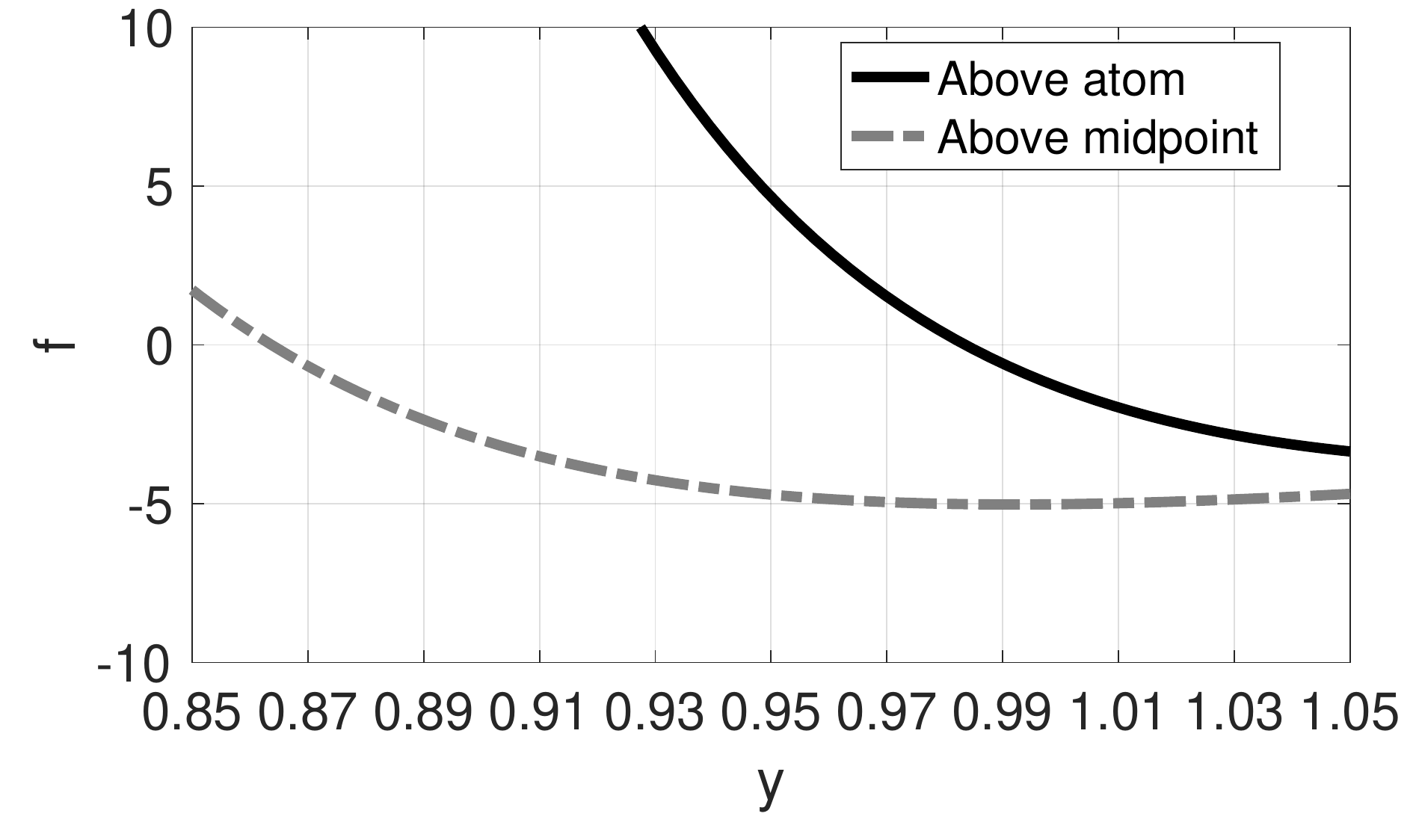}
\caption{{Van der Waals force on an atom of the deformable chain when this atom is the distance $y$ above the midpoint between two adjacent atoms of the rigid chain (dash-dotted line), or the distance $y$ directly above an atom of the rigid chain (solid line). The atom is in equilibrium when $f=0$.}}
  \label{feqd}
\end{figure}

{The amplitude of the wrinkles is determined by how close an atom
  on the deformable lattice wants to be to the rigid chain when it is
  in registry (i.e, it is positioned above the midpoint between two
  atoms of the rigid chain) versus when it is out of registry (i.e.,
  it is positioned exactly above one of the atoms of the rigid
  chain). As we can see in Figure \ref{feqd}, the equilibrium distance for an atom in registry and out of registry are, respectively, $\sim0.86$ and $\sim0.98$. The lower of these values corresponds to the distance between the chains when they are in registry in Figures \ref{f23}--\ref{f29}, while the higher value closely approximates the height of the wrinkles observed in the same figures, unless the amplitude is depressed due to a relatively large bending stiffness.}

We conclude this section by emphasizing that the wrinkling predicted by the
continuum model occurs because the interaction energy $G$ retains
information about the discrete character of the substrate.  Recall
that, in the nondimensional variables, $h_{1}$ is the
spacing between atoms on the rigid chain and $\sigma$ is the
equilibrium separation of the Lennard-Jones atom-to-atom potential.
Because of the fast decay of this potential, the equilibrium
separation essentially determines the range of the van der Waals
interaction between atoms.
If $h_{1}$ is close to $\sigma$, an atom on the deformable chain
interacts with only the one or two nearby atoms, both to the left and to the
right on the rigid chain.  The strength of the interaction varies
significantly in the regions above the `gaps' between atoms on the
rigid chain, which creates deep potential wells.  As explained above,
these potential wells along with the mismatch drive the formation of
isolated wrinkles.
Conversely, when $h_{1}$ is much smaller then $\sigma$, an atom
on the deformable chain interacts with a relatively large number of
nearby atoms on the rigid chain and the gaps between these nearby
atoms are relatively small.  Hence the atom 
essentially feels the average of this interaction, which is like a
continuum approximation in which the strength of the interaction
depends only on the vertical distance from the line of fixed atoms.
One would not expect the formation of isolated wrinkles in this case.

\begin{figure}[h!]
\centering
\includegraphics[width=0.5\linewidth]
{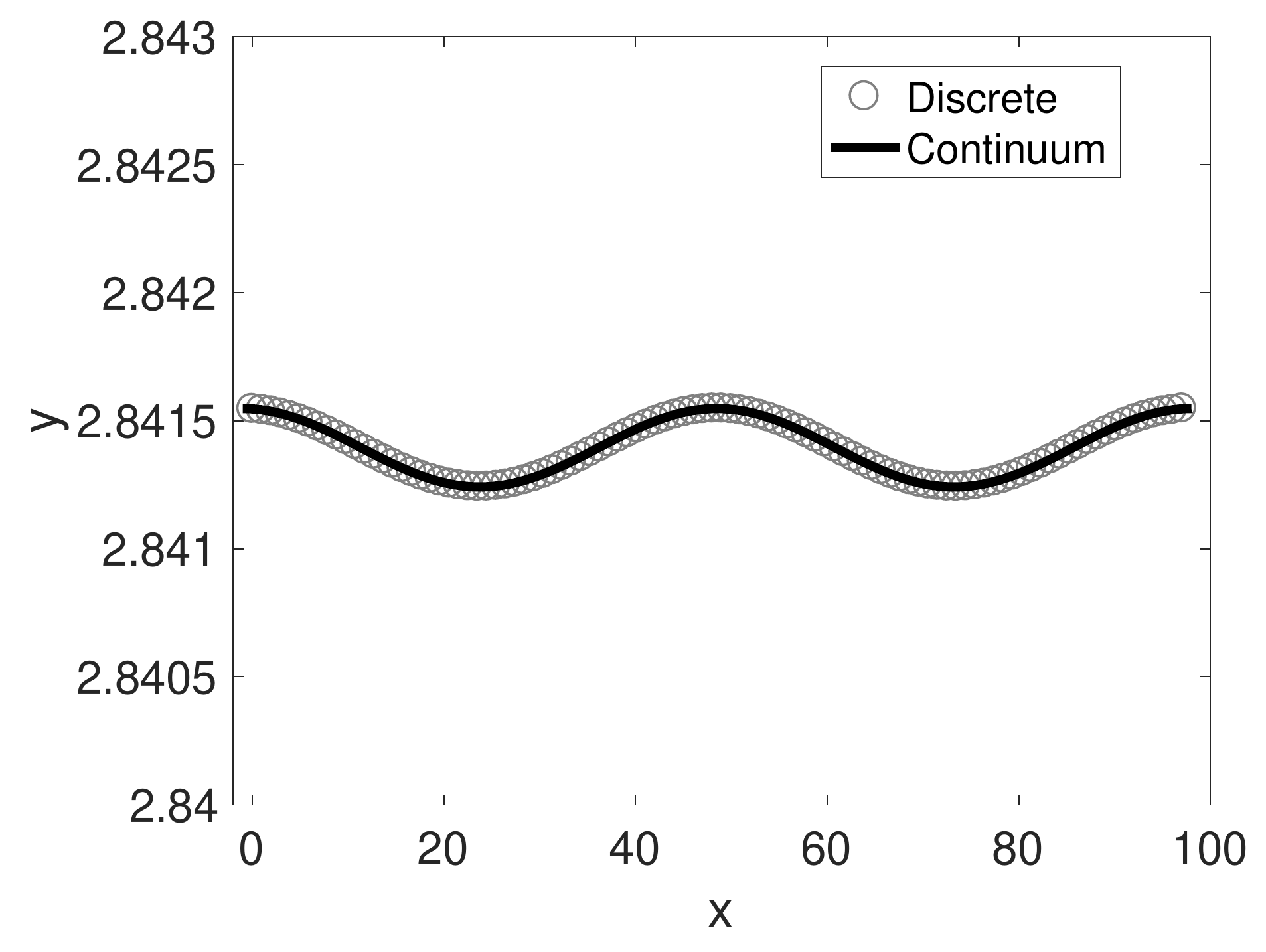}
\caption{{Discrete and continuum simulation results with $\sigma=3.0$, $\omega=1.0$, $h_{1}=1.01$, $h_{2}=.99$, $k_s=10.0$, and $k_b=100.0$. The discrete simulation involves $99$ atoms on the rigid chain and $101$ atoms on the deformable chain per period. Note that the vertical scale is
approximately 100 times smaller than the vertical scale in Figures~\ref{f23}--\ref{f29}.}}
  \label{f30}
\end{figure}

To support these observations, we note that
for the solutions shown in Figures~\ref{f23}---\ref{f29}, $h_{1}=\sigma=1.0$.  We contrast the isolated wrinkles observed in
those cases with the discrete and continuum solutions shown
in Figure~\ref{f30}.
For the discrete solution in Figure~\ref{f30}, $h_{1}=1.01$ and
$\sigma=3.0$ and the corresponding configuration is
approximately sinusoidal.  Furthermore, the amplitude of the solution
is on the order of $10^{-4}$, while the height of the wrinkles in
Figures~\ref{f23}---\ref{f29} is on the order of $10^{-2}$.  {The continuum solution for parameters
corresponding to those used in the discrete simulation shows an
excellent match with the discrete solution.}

\section{Conclusions}\label{sconc}

{We have applied an upscaling procedure to develop a mesoscopic
  continuum model of the cross-section of a graphene sheet interacting
  with a rigid crystalline substrate. Without making any a priori
  assumptions on the structure of the continuum model, we established
  that the elastic part of the model resembles the F\"oppl--von K\'arm\'an theory of thin
  rods but without thickness as that notion is meaningless for the 2D
  materials (or their 1D reduction). Our approach allows us to
  identify the potential function that keeps track of microscopic
  registry effects on the mesoscale and also determines the correct orders of magnitude of various contributions to the continuum energy in terms of the small parameter $\varepsilon$. The main novelty of this work is in the development of the continuum contribution for the weak interaction between the deformable and rigid chains.  Although a continuum description, this energy retains discrete information about any mismatch between the chains.}

{Numerical simulations demonstrate that the predictions of the
  mesoscopic model are generally in close correspondence with the
  configurations obtained via a  discrete molecular dynamics
  approach. In both cases, relaxation of slightly mismatched chains
  resulted in formation of wrinkles, where flat, approximately uniform
  domains, are separated by wrinkles characterized by relatively large
  out-of-plane deformations. Fixing the strength of the Lennard-Jones
  potential, and using a flat, undeformed configuration as the initial
  condition, we considered a number of different parameter
  regimes. When the elastic constants corresponding to stretching and
  bending are of order $1$, the wrinkles contain a small number of
  atoms and the quality of the continuum approximation near the top of
  the wrinkle is poor, while the approximation is accurate both near
  the base of the wrinkle and in the flat regions of registry. In the
  case when the stretching constant is of order
  $>10$ and is larger than the bending constant, which is of practical importance,  a solution of the
  continuum model approximates well the corresponding solution of the
  discrete model, even near the top of the wrinkle.  This improved
  approximation occurs because there are more atoms per wrinkle and
  because the solution exhibits relatively smaller gradients for a
  given $\eps$.}
  
  {Compared to the number of atoms on the rigid chain, there can
  be more or fewer atoms on the deformable chain. The number of
  wrinkles is then equal to the difference between the number of atoms on the chains, with each wrinkle compensating for exactly one extra
  atom or vacancy. The locations of the wrinkles correspond to the
  regions of the largest misfit in the moir\'e pattern for a given
  system of incommensurate deformable and rigid chains.   }
  
 {Unlike what is seen in
  analogous thin film/substrate systems with misfit strains, here the wrinkling is
  observed not only when the deformable chain is compressed but also when it is stretched. This is true because the equilibrium distance between an atom of the deformable chain and the rigid chain is strictly larger in the region of misregistry than in the region of registry. The difference between these equilibrium distances determines the amplitude of a wrinkle.}

{Periodic boundary conditions were imposed in this work primarily
  for simplicity, and other boundary conditions can be considered. The
  model can also be extended in a standard way to include applied body
  forces. In this more general setting it would be possible to consider a problem in which the registry effects can be combined with a standard compressive misfit between the deformable and rigid chains. It would be of interest to understand how the wrinkling pattern would respond to the applied loads and whether different types of singular regions---e.g., wrinkles and delamination blisters---may coexist in the same system.}

{Continuum modeling that retains discrete registry effects is important both for solving computational problems more efficiently and for allowing theoretical insight into mesoscopic pattern formation in graphene and other 2D materials. Here one of the principal advantages is the ability to fully utilize the powerful tools provided by the theory of differential equations and variational calculus. The similarity between the structure of the continuum model that was derived in this paper and the existing models of a thin film on an adhesive substrate should allow for an extension of the known results for these macroscopic systems to 2D materials. For example, the formation of dips in the vicinity of the wrinkles, which was discussed in the previous section, may likely be explained using the techniques developed in \cite{PhysRevLett.107.044301}. Continuum modeling may also facilitate the study of the influence of atomic relaxation of slightly mismatched graphene/substrate lattices on electronic properties of the system \cite{van2015relaxation}.}

{In addition to the analysis of the one-dimensional model, future work will include a rigorous verification of our conjecture that both the discrete and continuum models converge to the same asymptotic limit in the appropriate sense as $\varepsilon\to0$. Here the limit has to be understood within the framework of $\Gamma$-convergence \cite{braides2002gamma}. This analysis should justify the number of terms that was retained in the expansion of the discrete problem in order to obtain the continuum problem. Further, we plan to develop an extension of the derivation presented in this paper to a full model for a 2D material consisting of two slightly mismatched, interacting lattices.}

\bigskip
\textbf{Acknowledgment}:
This work was supported by the National Science Foundation grant DMS-1615952.

\bibliographystyle{ieeetr}
\bibliography{gamma-converg-references}	

\begin{thebibliography}{10}

\bibitem{braun2013frenkel}
O.~Braun and Y.~Kivshar, {\em The Frenkel-Kontorova Model: Concepts, Methods,
  and Applications}.
\newblock Theoretical and Mathematical Physics, Springer Berlin Heidelberg,
  2010.

\bibitem{van2015relaxation}
M.~M. van Wijk, A.~Schuring, M.~I. Katsnelson, and A.~Fasolino, ``Relaxation of
  moir{\'e} patterns for slightly misaligned identical lattices: graphene on
  graphite,'' {\em 2D Materials}, vol.~2, no.~3, p.~034010, 2015.

\bibitem{van2014moire}
M.~M. van Wijk, A.~Schuring, M.~I. Katsnelson, and A.~Fasolino, ``Moire
  patterns as a probe of interplanar interactions for graphene on h-{BN},''
  {\em Physical Review Letters}, vol.~113, no.~13, p.~135504, 2014.

\bibitem{PhysRevE.85.066115}
B.~Davidovitch, R.~D. Schroll, and E.~Cerda, ``Nonperturbative model for
  wrinkling in highly bendable sheets,'' {\em Phys. Rev. E}, vol.~85,
  p.~066115, Jun 2012.

\bibitem{Kohn2013}
R.~V. Kohn and H.-M. Nguyen, ``Analysis of a compressed thin film bonded to a
  compliant substrate: The energy scaling law,'' {\em Journal of Nonlinear
  Science}, vol.~23, no.~3, pp.~343--362, 2013.

\bibitem{wang2014phase}
Q.~Wang and X.~Zhao, ``Phase diagrams of instabilities in compressed
  film-substrate systems,'' {\em Journal of applied mechanics}, vol.~81, no.~5,
  p.~051004, 2014.

\bibitem{PhysRevLett.72.3570}
J.~Tersoff and F.~K. LeGoues, ``Competing relaxation mechanisms in strained
  layers,'' {\em Phys. Rev. Lett.}, vol.~72, pp.~3570--3573, May 1994.

\bibitem{CPA:CPA21643}
P.~Bella and R.~V. Kohn, ``Coarsening of folds in hanging drapes,'' {\em
  Communications on Pure and Applied Mathematics}, vol.~70, no.~5,
  pp.~978--1021, 2017.

\bibitem{PhysRevLett.107.044301}
T.~J.~W. Wagner and D.~Vella, ``Floating carpets and the delamination of
  elastic sheets,'' {\em Phys. Rev. Lett.}, vol.~107, p.~044301, Jul 2011.

\bibitem{PhysRevLett.105.038302}
J.~Huang, B.~Davidovitch, C.~D. Santangelo, T.~P. Russell, and N.~Menon,
  ``Smooth cascade of wrinkles at the edge of a floating elastic film,'' {\em
  Phys. Rev. Lett.}, vol.~105, p.~038302, Jul 2010.

\bibitem{CPA:CPA21471}
P.~Bella and R.~V. Kohn, ``Wrinkles as the result of compressive stresses in an
  annular thin film,'' {\em Communications on Pure and Applied Mathematics},
  vol.~67, no.~5, pp.~693--747, 2014.

\bibitem{golovaty2008continuum}
D.~Golovaty and S.~Talbott, ``Continuum model of polygonization of carbon
  nanotubes,'' {\em Physical Review B}, vol.~77, no.~8, p.~081406, 2008.

\bibitem{cazeaux2017analysis}
P.~Cazeaux, M.~Luskin, and E.~B. Tadmor, ``Analysis of rippling in
  incommensurate one-dimensional coupled chains,'' {\em Multiscale Modeling \&
  Simulation}, vol.~15, no.~1, pp.~56--73, 2017.

\bibitem{wilber2007continuum}
J.~P. Wilber, C.~B. Clemons, G.~W. Young, A.~Buldum, and D.~D. Quinn,
  ``Continuum and atomistic modeling of interacting graphene layers,'' {\em
  Physical Review B}, vol.~75, no.~4, p.~045418, 2007.

\bibitem{blanc_lebris_lions_2007}
X.~Blanc, C.~Le~Bris, and P.-L. Lions, ``Atomistic to continuum limits for
  computational materials science,'' {\em ESAIM: Mathematical Modelling and
  Numerical Analysis}, vol.~41, no.~2, p.~391–426, 2007.

\bibitem{braides2002gamma}
A.~Braides, {\em Gamma-convergence for Beginners}, vol.~22.
\newblock Clarendon Press, 2002.

\bibitem{braides2008asymptotic}
A.~Braides and L.~Truskinovsky, ``Asymptotic expansions by
  {$\Gamma$}-convergence,'' {\em Continuum Mechanics and Thermodynamics},
  vol.~20, no.~1, pp.~21--62, 2008.

\bibitem{lu2009elastic}
Q.~Lu, M.~Arroyo, and R.~Huang, ``Elastic bending modulus of monolayer
  graphene,'' {\em Journal of Physics D: Applied Physics}, vol.~42, no.~10,
  p.~102002, 2009.

\bibitem{lee2008measurement}
C.~Lee, X.~Wei, J.~W. Kysar, and J.~Hone, ``Measurement of the elastic
  properties and intrinsic strength of monolayer graphene,'' {\em science},
  vol.~321, no.~5887, pp.~385--388, 2008.

\end{thebibliography}

\end{document}